\documentclass[reqno]{amsart} 

\usepackage{amsmath}
\usepackage{amsfonts}
\usepackage{amssymb}
\usepackage{enumerate}
\usepackage{graphicx}

\DeclareGraphicsExtensions{.ps,.eps,.eps.gz}

\theoremstyle{plain}
\newtheorem{thm}{Theorem}[section]{\bf}{\it}
\newtheorem{prop}[thm]{Proposition}{\bf}{\it}
\newtheorem{lemma}[thm]{Lemma}{\bf}{\it}
{\bf}{\it}
{\bf}{\it}
{\bf}{\it}

{\bf}{\rm}
{\it}{\rm}
\newtheorem{remark}[thm]{Remark}{\it}{\rm}
\newtheorem{obs}[thm]{Observation}{\it}{\rm}

\newenvironment{pf*}[1]{\par\medskip\noindent\textit{#1}\,:}{\hspace*{\fill}\qed\medskip\par\noindent}

\numberwithin{equation}{section}
\newcommand{\supp}{\operatorname{supp}}

\newcommand{\Vol}{{\operatorname{Vol}}}
\newcommand{\Res}{{\operatorname{Res}}}

\newcommand{\Tr}{{\operatorname{Tr}}}
\newcommand{\N}{{\mathbb N}}
\newcommand{\R}{{\mathbb R}}
\newcommand{\C}{{\mathbb C}}

\begin{document}

\title{The large - $Z$ behaviour of pseudo-relativistic
atoms}

\author{Thomas \O stergaard S\o rensen}

\address{Mathematisches Institut,
         Universit\"at M\"unchen,
         Theresienstra\ss e 39,
         D-80333 {Munich},
         Germany}
\address[On leave from]
         {Department of 
         Mathematical Sciences,
         Aalborg University,
         Fredrik Bajers Vej 7G,
         DK-9220 Aalborg East,
         Denmark}
\address[]{}

\email{sorensen@mathematik.uni-muenchen.de}

\renewcommand{\datename}{}
\date{\today}
\thanks{\copyright\ 2004 by the 
       author. This article may be reproduced in its entirety for
       non-commercial purposes.}

\renewcommand{\subjclassname}{\textup{2000} Mathematics Subject Classification}

\begin{abstract} 
In this paper we study the large - $Z$ behaviour of the ground 
state energy of atoms with electrons having
relativistic kinetic energy
$\sqrt{p^{2}c^{2}+m^{2}c^{4}}-mc^{2}$. 
We prove that to leading order
in $Z$ the energy is the same as in the non-relativistic case, given
by (non-relativistic) Thomas-Fermi theory. For the problem to make
sense, we keep the 
product $Z\alpha$ fixed (here $\alpha$ is Sommerfeld's fine structure
constant), and smaller than, or equal to, $2/\pi$, which means that as
$Z$ tends to 
infinity, $\alpha$ tends to zero.
\end{abstract}

\maketitle

\section{Introduction and results}
\label{intro}

As a model for a relativistic atom with nuclear charge $Z$ and $N$
electrons, we consider the operator
\begin{eqnarray*}
H_{\text{\it{rel}}}= \sum_{i=1}^{N}\left\{\sqrt{
    -\alpha^{-2}\Delta_{i}+\alpha^{-4}} - \alpha^{-2} -
   \frac{Z}{|x_{i}|}\right\}
  + \sum_{1\leq i<j \leq N}\frac{1}{|x_{i}-x_{j}|}.
\end{eqnarray*}
Here, \(x_{i}\in\mathbb R^{3}\) is the coordinate of the \(i\)'th
electron, \(\Delta_{i}\) is the Laplacian with respect to \(x_i\), and
 \(\alpha\) is Sommerfeld's fine structure constant (the physical
value of \(\alpha\) is approximately 1/137.037). 
This is the expression one obtains using
\(\sqrt{p^2c^2+m^2c^4}-mc^2\) for the kinetic energy of the electrons
(and making the substitution \(p\to -i\hbar\nabla\)),
measuring energies (\(H_{\text{\it{rel}}}\)) in units of Ryd\-berg, and
lengths (the \(x_{i}\)'s) in units of the
Bohr radius.

This model has been much studied over the past thirty
years. Stability in the case $N=1$ was proved independently 
by Herbst \cite{He77} and Weder \cite{We75}. The `Stability of Matter'
for the model was first proved by Conlon~\cite{Co84}, later by
Fefferman and de la Llave~\cite{FeLla86}, and also by Lieb and
Yau~\cite{LiYau88}; see the latter for an overview. A non-exhaustive
list of other works
on this model is \cite{LewSieVu,ZhiVu1,ZhiVu2,ZhiVu3,SieBenSto}.

It is well-known that the operator $H_{\text{\it{rel}}}$ is bounded
from below on $C_{0}^{\infty}(\R^{3N})$ if, and only if,
$Z\alpha\leq\frac{2}{\pi}$. Only in this case is the atom stable; and
we define the operator $H_{\text{\it{rel}}}$ as a self-adjoint,
unbounded operator by Friedrichs-extending
this semi-bounded operator. To study the energy of large
atoms, one would normally then consider the limit as $Z\to\infty$ of
the infimum of the spectrum of this operator. However, due to the upper bound
on $Z$ resulting from the restriction $Z\alpha\leq\frac{2}{\pi}$,
this is not possible here. To overcome this problem, we 
consider
\begin{eqnarray*}
H_{\text{\it{rel}}}=
 \alpha^{-1}\left\{ \sum_{i=1}^{N}\left\{\sqrt{-
 \Delta_{i}+\alpha^{-2}} - \alpha^{-1} -\frac{\delta}{|x_{i}|}\right\}
 +\sum_{1\leq i<j \leq N}\frac{\alpha}{|x_{i}-x_{j}|} \right\}
\end{eqnarray*}
where $\delta=Z\alpha$ is held {\it fixed}. This ensures that as
$Z\to\infty$, and therefore $\alpha\to0$, the operator 
$H_{\text{\it{rel}}}$ remains well-defined---as long as
$0\leq\delta\leq\frac{2}{\pi}$. Also,
we shall keep $\lambda\equiv N/Z$ fixed. 
The energy of the atom is then defined
as 
\begin{equation*}
   E_N(Z,\delta):=\inf \sigma_{{\mathcal H}_F}
   (H_{\text{\it{rel}}})\ ,
\end{equation*}
where the spectrum of $H_{\text{\it{rel}}}$ is calculated on
${\mathcal H}_F=\bigwedge^N L^2(\R^3,\C^q)$, the Fermionic
Hilbert space, describing $N$ Fermions, each with $q$ possible spin
states. We will take $q=2$ from now on (but this is no
restriction). We note that since 
(the extension of) $H_{\text{\it{rel}}}$ is self-adjoint and
bounded from below, we have the Rayleigh-Ritz principle:
If $\mathcal{C}$ is a form core for the corresponding quadratic form, then
\begin{equation*}
 \inf \sigma_{{\mathcal H}_F}
 (H_{\text{\it{rel}}})
 =\inf_{\{\psi\in\mathcal{C}\,|\, \|\psi\|=1\}}
 \langle\psi,H_{\text{\it{rel}}}\,\psi\rangle\ .
\end{equation*}
Our main result is the following:
\begin{thm}
\label{thm:TF}
Let $\delta\in(0,2/\pi]$ and \(\lambda>0\) be fixed and
let $H_{\text{\it{rel}}}$ and $E_{N=\lambda Z}(Z,\delta)$
be as above. Then
\begin{equation}
\label{eq:TF}
  E_{\lambda Z}(Z,\delta)= {}-C_{\text{TF}}(\lambda)Z^{7/3} +
  \text{o}(Z^{7/3})\quad, \quad Z\to\infty, 
\end{equation}
where $-C_{\text{TF}}(\lambda)Z^{7/3}$ is the (non-relativistic) Thomas-Fermi energy of the atom.
\end{thm}
This shows that, to leading order, the ground-state energy of a relativistic
atom is given by the (non-relativistic) semi-classical Thomas-Fermi energy approximation, as
it is for the non-relativistic atom (note that the case
$\delta=\frac{2}{\pi}$ is included). (In the non-relativistic case this
was first proved 
by Lieb and Simon \cite{LiSi77}; see also Lieb \cite{Li81b}.) This
expresses the fact that for large atoms  the majority of the
electrons are non-relativistic.

The second term in the expansion~\eqref{eq:TF} will be studied in a
forthcoming paper~\cite{RelScott}.

The proof of Theorem~\ref{thm:TF} will be by finding upper and lower bounds on 
$E_{\lambda Z}(Z,\delta)$. 
Note that the relativistic kinetic energy is 
always lower than the 
non-relativistic one:
\begin{equation}
  \sqrt{\alpha^{-2}q^{2}+\alpha^{-4}}-\alpha^{-2}=
  \alpha^{-2}\left(\sqrt{1+(\alpha q)^2}-1\right) \leq
  \frac{q^2}{2} .
\end{equation}
(Note: since we will later make Taylor expansions of 
the square  root in the relativistic kinetic energy, we will have to 
insist on the non-relativistic kinetic energy being $-\Delta/2$).
This means that all upper bounds derived earlier \cite{LiSi77,Li81b}
for the non-relativistic  
operator
\begin{equation*}
  H_{cl}= \sum_{i=1}^{N}\left\{ \frac{p_{i}^2}{2}-
   \frac{Z}{|x_{i}|}\right\}
  + \sum_{1\leq i<j \leq N}\frac{1}{|x_{i}-y_{j}|}
\end{equation*}
will also be upper bounds for \(H_{\text{\it rel}}\); in particular, to
prove Theorem~\ref{thm:TF}, we need 
only derive a lower bound.

\section{Organisation of the paper}
\label{sketch}

We start in Section~\ref{one_par} by reducing the $N$-body operator
\(H_{\text{\it rel}}\) to
a one-particle one; having done that, we only need to 
consider wave functions given as 
Slater-deter\-minants when trying to minimise the energy.
To proceed, we need to localise the kinetic energy. To do so, we
use (in Section~\ref{loc}) an analogue of the IMS
Localisation Formula for the Schr\"odinger 
operator, see~\cite[p.27]{CFKS}. This formula has already been developed
by Lieb and Yau in~\cite{LiYau88} for both the operator 
$\sqrt{-\Delta+\alpha^{-2}}-\alpha^{-1}$ and
the hyper-relativistic kinetic energy $|p|$. This is essentially done
by finding the integral kernels of these operators. For 
$\sqrt{-\Delta+\alpha^{-2}}-\alpha^{-1}$, this involves
the modified Bessel function $K_{2}$, and the derivation of the formula
and of needed properties of $K_{2}$ are carried out in
Appendix~\ref{appendixA}.
The localisation error, given by a bounded operator $L^{(\alpha)}$
expressed as an integral operator involving $K_{2}$, is then estimated
(in Section~\ref{loc_error}).
Estimating the error is rather technical (calculative) and involves 
localisation of the operator and the above mentioned properties of 
$K_{2}$. Some
of the localised terms are estimated with the localised energy itself
(Sections~\ref{inner} and \ref{inter}).

Coming to the localised energy, we have to estimate the kinetic energy 
close to the nucleus. Since this is the high-energy region, this is
where the electrons are  
relativistic, and so this term
should be of lower order, since, to leading order, there should be
no relativistic contribution to the energy. As the relativistic kinetic
energy is asymptotically linear in $p$ in the high-energy 
region---as opposed to the classical one which 
is quadratic---the singularity in the potential causes substantially
more trouble. This problem is solved (in Section~\ref{inner}) by a
clever choice of parameters 
in an estimate by Lieb and Yau in~\cite{LiYau88} on the sum of the 
eigenvalues of the energy in a ball 
around the nucleus. This also determines the scale on which one can localise
close to the nucleus. A part of two of the localised terms of the operator
$L^{(\alpha)}$ is estimated along with this term.

In the outer region, one uses (in Section~\ref{outer}) essentially the same
idea as Lieb did in the classical case, see~\cite{Li81a}, to re-find
the desired phase space integral, which is to give the semi-classical 
Thomas-Fermi energy. This involves introducing coherent states and 
estimating the error by doing so. The formulae for the relativistic
case were developed in~\cite{LiYau87}, but the error obtained there is too
rough for our purposes. We therefore develop (in
Appendix~\ref{appendixB}) a better estimate by a more 
careful analysis.
In order to make all this work, one need the coherent state to be supported
further out than the initial cut-off around the nucleus. To get this, 
an intermediary zone is introduced (also in Section~\ref{loc}) by an additional
cut-off. The energy in this shell is 
estimated (in Section~\ref{inter}) by a generalised version of the Lieb-Thirring inequality,
proved by Daubechies in \cite{Dau83}. Also the other part of the 
previously mentioned two terms of the localised
operator $L^{(\alpha)}$ is estimated in this way.

Finally we relate (in Section~\ref{outer}) the energy in the outer region to 
the Thomas-Fermi energy from the classical 
(that is, the Schr\"odinger) case. In this region, the
kinetic energy is small, and using
the specific scaling property of Thomas-Fermi theory
allows one to make the change from the relativistic
energy $\sqrt{-\alpha^{-2}\Delta+\alpha^{-4}}-\alpha^{-2}$ to the
non-relativistic one, $-\Delta/2$, getting errors of the desired order.

\section{Reduction to a one-particle problem}
\label{one_par}

We will use the notation
\begin{align}
 \label{def:H}
 H=\alpha H_{\text{\it{rel}}}
  =
 \sum_{i=1}^{N}\Big\{\sqrt{-\Delta_{i}+\alpha^{-2}}-\alpha^{-1}-
 \frac{\delta}{|x _{i}|}\Big\}+\sum_{1\leq i<j\leq
 N}\frac{\alpha}{|x_{i}-x_{j}|}.
\end{align}
Recall that $\delta=Z\alpha$ is {\it fixed} and that the ground state
energy of $H_{\text{\it{rel}}}$ is to be proven to be of leading order
$Z^{7/3}$. Since we wish to consider $\alpha$ as the free parameter,
the relevant order of all error terms will be $o(\alpha^{-4/3})$.
Also, we
will denote the operator $\sqrt{-\Delta+\alpha^{-2}}$ by
$\sqrt{p^{2}+\alpha^{-2}}$, and so
$T(p)=\sqrt{p^{2}+\alpha^{-2}}-\alpha^{-1}$
will be the kinetic energy.

We start by reducing the problem from an $N$-particle problem
to a one-particle one. This is done by using an inequality on the
electron-electron interaction $\sum_{i<j}|x_{i}-x_{j}|^{-1}$, which 
will reduce this to a one-particle potential.

Choose a spherically symmetric function $g\in
C_{0}^{\infty}(\R^{3})$, non-negative, supported in the unit ball
$B(0,1)$ of $\R^{3}$, and such that
$\int\!g(x)^{2}\,d^{3}\!x=1$.
Let $\phi(x)=g(x)^{2}$ and let
for $a>0$ ($a$ to be chosen later),
$\phi_{a}(x)=a^{-3}\phi(x/a)$, so that $\int\!\phi_{a}(x)\,d^{3}\!x=1$. 
Then for all $\rho: \R^{3}\to\R$ we have:
\begin{align*}
  &\sum_{1\leq i<j\leq N}\frac{1}{|x_{i}-x_{j}|} \geq
  \sum_{1\leq i<j\leq N}\iint\frac{\phi_{a}(x-x_{i})\phi_{a}(y-x_{j})}{|x-y|}\,
  d^{3}\!x\,d^{3}\!y\\
  &\ \ =
  \frac{1}{2}\sum_{i,j=1}^{N}\iint\frac{\phi_{a}(x-x_{i})\phi_{a}(y-x_{j})}
  {|x-y|}\,d^{3}\!x\,d^{3}\!y 
  - \frac{1}{2}\,N\iint\frac{\phi_{a}(x)\phi_{a}(y)}
  {|x-y|}\,d^{3}\!x\,d^{3}\!y \\
  &\ \ =
  \sum_{i=1}^{N}\iint\frac{\rho(y)\phi_{a}(x-x_{i})}
  {|x-y|}\,d^{3}\!x\,d^{3}\!y
  -\frac{1}{2}\iint\frac{\rho(x)\rho(y)}{|x-y|}\,d^{3}\!x\,d^{3}\!y
  - c(\phi)Na^{-1}\\
  &\qquad+\frac{1}{2}\iint\frac{\Big(\sum_{i}\phi_{a}(x-x_{i})
  -\rho(x)\Big)
  \Big(\sum_{j}\phi_{a}(y-x_{j})-\rho(y)\Big)}{|x-y|}\,d^{3}\!x\,d^{3}\!y\\
  &\ \ \geq \sum_{i=1}^{N}\iint\frac{\rho(y)\phi_{a}(x-x_{i})}{|x-y|}
  \,d^{3}\!x\,d^{3}\!y
  -\frac{1}{2}\iint\frac{\rho(x)\rho(y)}{|x-y|}\,d^{3}\!x\,d^{3}\!y
    - c(\phi)Na^{-1}.
\end{align*}
In the last inequality we used that $|x-y|^{-1}$ is of positive type
(a positive kernel) since 
\begin{equation*}
  \iint \frac{\overline{f(x)}f(y)}{|x-y|}\,d^{3}\!x\,d^{3}\!y
  =4\pi\int\frac{|\hat f(p)|^{2}}{|p|^{2}}\,d^{3}\!p.
\end{equation*}
The constant $c(\phi)$ is independent of
$a$:
\begin{equation*}
  c(\phi)=\frac{1}{2}\iint\frac{\phi(x)\phi(y)}{|x-y|}\,d^{3}\!x\,d^{3}\!y
  =2\pi\int\frac{|\hat \phi(p)|^{2}}{|p|^{2}}\,d^{3}\!p.
\end{equation*}
Noting that (using the spherical symmetry of $\phi_{a}$)
\begin{align*}
  &\iint\frac{\rho(y)\phi_{a}(x-x_{i})}{|x-y|}\,d^{3}\!x\,d^{3}\!y
  =\iint\frac{\rho(y)\phi_{a}(z)}{|z-(x_{i}-y)|}\,d^{3}\!z\,d^{3}\!y\\
  &=\int\rho(y) \big(\phi_{a}*|\cdot|^{-1}\big)(x_{i}-y)\,d^{3}\!y
  =\big(\rho*\phi_{a}*|\cdot|^{-1}\big)(x_{i})
  \equiv\rho*\phi_{a}*|x_{i}|^{-1},
\end{align*}
we get the operator inequality (see \eqref{def:H} for \(H\)):
\begin{align}
 \label{eq:one_part_op}
 H &\geq\sum_{i=1}^{N}\Big\{\sqrt{p_{i}^{2}+\alpha^{-2}}-\alpha^{-1}
  -\frac{\delta}{|x_{i}|} 
 +\alpha\,\rho*\phi_{a}*|x_{i}|^{-1}\Big\}\\
 &\qquad\qquad-\frac{\alpha}{2}\iint\frac{\rho(x)\rho(y)}{|x-y|}
 \,d^{3}\!x\,d^{3}\!y
 -\alpha\,c(\phi)Na^{-1}\notag.
\end{align}

Having reduced the $N$-body operator $H$ to a one-body operator, 
we only need to consider Slater-determinants when trying to
minimise the energy. That is, when considering $\langle\psi,
H\psi\rangle$ we need only consider those $\psi\in{\mathcal H}_{F}$ which
are given by 
\begin{equation*}
\psi(x_{1},\dots,x_{N})=\frac{1}{\sqrt{N!}}\det(m_{i}(x_{j})),
\end{equation*}
where $m_{i}\in L^2(\R^3), i=1,\dots,N$, are orthonormal. Note also
that since $C_{0}^{\infty}(\R^{3})$ is a core for the operator
$\sqrt{p^{2}+\alpha^{-2}}-\alpha^{-1}-\delta/|x|, \delta\in[0,2/\pi]$ (see
Herbst~\cite{He77}), we need only
consider $m_{i}$'s in this space.
Then, as soon as $h$ is a one-particle operator acting on $L^2(\R^3)$,
we have that
\begin{equation*}
  \langle\psi, \sum_{i=1}^{N}h_{i} \psi\rangle = 
  \sum_{i=1}^{N}(m_{i},h\,m_{i}).
\end{equation*}
Here, $\langle\ ,\ \rangle$ and $(\ ,\ )$ denote inner products
in $L^{2}(\R^{3N})$, respectively $L^{2}(\R^{3})$, both linear in the
second variable,  and $h_{i}$ is $h$
acting on the variable $x_{i}$ of $\psi$. Also, we will use
$\|\cdot\|_{p}$ for the norm in $L^{p}(\R^{3})$.

\section{Localisation of the kinetic energy}
\label{loc}

\begin{figure}[htbp]
  \begin{center}
     \includegraphics[width=4.5in, height=1in]{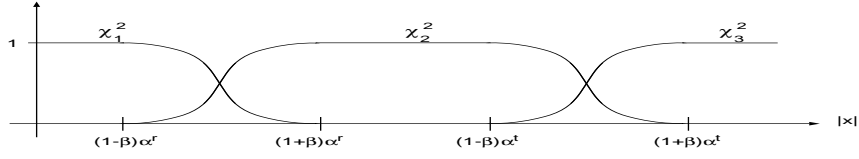}
    \caption{\sl The partition of unity.}
  \label{figurnr1}
  \end{center}
\end{figure}

In order to treat the one-body operator in~(\ref{eq:one_part_op}) and
in particular the singularity in the Coulomb-potential---which causes 
considerably more trouble than in the non-relativistic case---we introduce, 
following Lieb and Yau~\cite{LiYau88}, a partition of unity (see 
also Cycon, Froese, Kirsch and Simon~\cite[Definition 3.1]{CFKS}).
For some $\beta\in(0,\frac{1}{2})$, let $\theta_{1}$ and  $\theta_{2}$
be monotone positive 
smooth functions on $\R_{+}$, $0\leq\theta_{i}\leq1$, such that 
\begin{equation*}
\begin{split}
  \theta_{1}(\xi) = \left\{ \begin{array}{ll}
                        1  & \text{if $\xi < 1-\beta$} \\
                        0  & \text{if $\xi > 1+\beta$} \\
                           \end{array}
                   \right.
   \qquad,\qquad
    \theta_{2}(\xi) = \left\{ \begin{array}{ll}
                        0  & \text{if $\xi < 1-\beta$}  \\
                        1  & \text{if $\xi > 1+\beta$} \\
                           \end{array}
                   \right.
,
\end{split}
\end{equation*}
and such that $\theta_{1}(\xi)^{2}+\theta_{2}(\xi)^{2}=1$
for all $\xi\in\R_{+}$. Now define, with
$8/9<r<1$ and $1/3<t<2/3$ (these choices of parameters are governed by the
later analysis), the following three functions, which (for $\alpha$ sufficiently
small) is a 
partition of unity
in $\R^{3}$:
\begin{align}
   \label{eq:chis}
   \chi_{1}(x) = \theta_{1}\Big(\frac{|x|}{\alpha^{r}}\Big),\quad
   \chi_{2}(x) = \theta_{1}\Big(\frac{|x|}{\alpha^{t}}\Big)\,
                  \theta_{2}\Big(\frac{|x|}{\alpha^{r}}\Big),\quad 
   \chi_{3}(x) = \theta_{2}\Big(\frac{|x|}{\alpha^{t}}\Big).
\end{align}
Then, at least for $\alpha$ sufficiently small, we have the picture in
Figure~\ref{figurnr1}.
\\
\noindent According to Lieb and Yau~\cite[Theorem 9; 
$\alpha^{-1}$ corresponds to $m$]{LiYau88} we have for 
$f\in C_{0}^{\infty}(\R^{3})$ that
\begin{eqnarray}
\label{eq:loc_formula}
  (f, \sqrt{p^2+\alpha^{-2}} f)=
  \sum_{j=1}^{3} 
  (f, \chi_{j}\sqrt{p^2+\alpha^{-2}}\chi_{j} f)
   -  (f, L^{(\alpha)} f)
\end{eqnarray}
where $L^{(\alpha)}$ is a bounded operator on $L^2(\R^3)$, given by the kernel
\begin{equation*}
L^{(\alpha)}(x,y) = \frac{\alpha^{-2}}{4\pi^2}\, 
\frac{K_2(\alpha^{-1}|x-y|)}{|x-y|^{2}}
\sum_{j=1}^{3}(\chi_{j}(x)-\chi_{j}(y))^{2}.
\end{equation*}
Here $K_2$ is a modified Bessel-function, defined  on $(0,\infty)$ by
\begin{equation*}
  K_2(t)=\frac{1}{2}\int_{0}^{\infty}x e^{-\frac{1}{2}t(x+x^{-1})}\, dx.
\end{equation*}
For completeness, we derive this in Appendix~\ref{appendixA}.

Using this we find, with $T(p)=\sqrt{p^2+\alpha^{-2}}
-\alpha^{-1}$, $V(x)=\delta/|x|$ and $\psi$ a Slater-determinant as
mentioned in the previous section, that
\begin{align}
  \label{eq:slater}
  \langle\psi,&\sum_{i=1}^{N}
  \big\{T(p_{i})-V(x_{i})+\alpha\,\rho*\phi_{a}*|x_{i}|^{-1}\big\}
  \psi\rangle\\
  &=\sum_{i=1}^{N}(m_{i}, \big\{T(p)-V(x)
  +\alpha\,\rho*\phi_{a}*|x|^{-1}\big\}m_{i})\nonumber\\
  &=\sum_{j=1}^{3}
  \sum_{i=1}^{N}(m_{i},\chi_{j}\big\{ T(p)-V(x)
  +\alpha\,\rho*\phi_{a}*|x|^{-1}\big\}
  \chi_{j}m_{i})-\sum_{i=1}^{N}(m_{i},L^{(\alpha)}m_{i}),
  \nonumber
\end{align}
since $\sum_{j=1}^{3}\chi_{j}(x)^{2}=1$ for all $x\in\R^{3}$ (and
$\alpha$ sufficiently small).

\section{The localisation error}
\label{loc_error}

We now estimate the error introduced by the localisation of the
kinetic energy carried out in the last section. This error
is given by a bounded operator $L^{(\alpha)}$, 
\begin{align*}
  &L^{(\alpha)}(x,y)
  =\sum_{j=1}^{3} L_{j}^{(\alpha)}(x,y)\quad,\quad
  L^{(\alpha)}_{j}(x,y)=\frac{\alpha^{-2}}{4\pi^{2}}\,
  \frac{K_{2}(\alpha^{-1}|x-y|)}{|x-y|^{2}}(\chi_{j}(x)-\chi_{j}(y))^{2}.
\end{align*}
As noted above, this expression is derived in Appendix~\ref{appendixA}.
We shall start by localising this operator, thereby splitting it it into
twelve terms (!) which we will then treat individually.
These terms are going to fall into groups though, and  the terms in each of 
these will be estimated  together by different means. Two of the
terms will be estimated in later sections, together with the
energies near the nucleus and in the intermediary zone, related
to respectively $\chi_{1}$ and $\chi_{2}$.

In this section, the scale of the inner cut-off will be called
$l$, that is, $l=\alpha^{r}$, $8/9<r<1$. Let $\chi_{-}$ be the 
characteristic function of the ball $B(0,2l)$ in $\R^{3}$ and 
$\chi_{+}$ that for the
complement of this ball. Then each $L_{j}^{(\alpha)}$, $j=1,2,3$, splits
into four terms:
\begin{align*}
  L_{j}^{(\alpha)}(x,y)&=\chi_{+}(x)L_{j}^{(\alpha)}(x,y)\chi_{+}(y)
                       +\chi_{+}(x)L_{j}^{(\alpha)}(x,y)\chi_{-}(y)\\
                       &\,+ \chi_{-}(x)L_{j}^{(\alpha)}(x,y)\chi_{+}(y)
                       +\chi_{-}(x)L_{j}^{(\alpha)}(x,y)\chi_{-}(y).
\end{align*}
The following lemma will eventually take care of six of these twelve terms:
\begin{lemma}
\label{lem:decay}
Let $l=\alpha^{r}$, $8/9<r<1$ and assume that, with $\gamma\equiv
1-\frac{b}{a}>0$, 
\begin{equation*}
 |x|> a l \ \text{ and } \  |y|<b l \  
  \text{ on }\ 
  \supp\ \chi_{+}(x)L_{j}^{(\alpha)}(x,y)\chi_{-}(y).
\end{equation*}
Then, for $f\in L^{2}(\R^{3})$, 
\begin{equation*}
  |(f,\chi_{+}L_{j}^{(\alpha)}\chi_{-}f)| \leq\rho(\alpha)
   \|f\|_{2}^{2}\,,
\end{equation*}
where $\rho(\alpha)=o(e^{-2\epsilon\alpha^{r-1}})$ as $\alpha\to 0$
for all $\epsilon$ such that $0<\epsilon<\gamma$. In particular,
$\rho(\alpha)=o(\alpha^{n})$ as $\alpha\to 0$ for all $n\in\N$.
\end{lemma}
\begin{remark}
Note that the result with \(x\) and \(y\) interchanged also holds.
\end{remark}
\begin{proof}
By assumption we have that 
\begin{equation*}
  |x-y|>\gamma|x| \ \text{ on }\ \supp\ \chi_{+}L_{j}^{(\alpha)}\chi_{-}.
\end{equation*}
Since both $|x|^{-2}$ and 
$K_{2}(\alpha^{-1}|x|)$ are decreasing in $|x|$ 
(the last is obvious from the definition of $K_{2}$), and 
since $(\chi_{j}(x)-\chi_{j}(y))^{2}\leq 1$,
we get that pointwise,
\begin{equation*}
 \chi_{+}(x)L_{j}^{(\alpha)}(x,y)\chi_{-}(y)
 \leq \chi_{+}(x)\frac{\alpha^{-2}}{4\pi^{2}}
  \frac{K_{2}(\alpha^{-1}\gamma|x|)}{(\gamma|x|)^{2}}
  \,\chi_{-}(y)
\end{equation*}
on  $\supp\ \chi_{+}L_{j}^{(\alpha)}\chi_{-}$. Therefore
\begin{align}
  \label{eq:loc1}
  |(f,\chi_{+}&L_{j}^{(\alpha)}\chi_{-}f)| \\
  &\leq 
  \left(\int |f(y)|\,\chi_{-}(y)\,d^{3}\!y\right) \left( 
  \frac{(\alpha\gamma)^{-2}}{4\pi^{2}}
  \int |f(x)| \,\chi_{+}(x) 
  \frac{K_{2}(\alpha^{-1}\gamma|x|)}{|x|^{2}}\,d^{3}\!x\right).
  \nonumber
\end{align}
We estimate both of these terms using the Cauchy-Schwartz
inequality. For the first we get
\begin{align}
 \label{eq:loc2}
 \int |f(y)|\,\chi_{-}(y)\,d^{3}\!y \leq \|f\|_{2}\,\|\chi_{-}\|_{2}
 =C l^{3/2}\|f\|_{2},
\end{align}
and for the second 
\begin{align}
  \label{eq:loc3}
 \int |f(x)| \,\chi_{+}(x)
  \frac{K_{2}(\alpha^{-1}\gamma|x|)}{|x|^{2}}\,d^{3}\!x 
  \leq \|f\|_{2}\, 
  \left(\int \Big(\chi_{+}(x)
  \frac{K_{2}(\alpha^{-1}\gamma|x|)}{|x|^{2}}\Big)^{2}
  \,d^{3}\!x\right)^{1/2}.
\end{align}
Using the estimate~\eqref{eq:K_{2}-est} in Appendix~\ref{appendixA} 
on $K_{2}$, we get the estimate
\begin{align*}
  \int \Big(&\chi_{+}(x)
  \frac{K_{2}(\alpha^{-1}\gamma|x|)}{|x|^{2}}\Big)^{2}\,d^{3}\!x \\ &\leq
  4\pi\int_{2l}^{\infty} \frac{16}{|x|^{4}}\frac{\pi e^{-2\alpha^{-1}\gamma|x|}}{2\alpha^{-1}
  \gamma|x|}
  \Big(1+(2\alpha^{-1}\gamma|x|)^{-1}
  +(2\alpha^{-1}\gamma|x|)^{-2}\Big)^{2}|x|^{2}\, d|x| \\
  &= 128\pi^{2}\alpha^{-1}\gamma\int_{4\gamma l\alpha^{-1}}
  ^{\infty}t^{-3}e^{-t}(1+\frac{1}{t}+\frac{1}{t^{2}})^{2}\, dt,
\end{align*}
where the last equality follows by the change of variables 
$t=2\gamma\alpha^{-1}|x|$.
Dominating $e^{-t}$ in the integrand by $e^{-4\gamma l\alpha^{-1}}$
and working out the resulting integral, we arrive at
(using~\eqref{eq:loc1}, \eqref{eq:loc2}, and \eqref{eq:loc3}; recall
that $l=\alpha^{r}$) 
\begin{equation*}
 |(f,  \chi_{+}L_{j}^{(\alpha)}\chi_{-}f)|
 \leq C\,\|f\|_{2}^{2}\, \alpha^{(3r-5)/2}e^{-2\gamma\alpha^{r-1}}
 \Big\{\dots\Big\}^{1/2}
\end{equation*}
where 
\begin{multline*}
 \Big\{\dots\Big\}^{1/2}=
 \Big\{\frac{1}{4}(4\gamma)^{-4}\alpha^{4(1-r)}+\frac{2}{5}(4\gamma)^{-5}
       \alpha^{5(1-r)} \\
       +\frac{1}{2}(4\gamma)^{-6}\alpha^{6(1-r)}+
       \frac{2}{7}(4\gamma)^{-7}\alpha^{7(1-r)}
       +\frac{1}{8}(4\gamma)^{-8}\alpha^{8(1-r)}\Big\}^{1/2}.
\end{multline*}  
Now, since $8/9<r<1$, this term tends to zero as $\alpha$ tends to 
zero. Also 
\begin{equation*}
  \alpha^{(3r-5)/2}e^{-2\gamma\alpha^{r-1}} 
  = o(e^{-2\epsilon\alpha^{r-1}})\ , \ \alpha\to0,
\end{equation*}
for all $\epsilon$ satisfying $0<\epsilon<\gamma$. This
proves the lemma.
\end{proof}
We now return to investigating the above mentioned twelve terms. 
Firstly, note
that two of these terms are actually zero:
\begin{align*}
  \chi_{+}(x)L_{1}^{(\alpha)}(x,y)\chi_{+}(y) &\equiv 0 \\
  \chi_{-}(x)L_{3}^{(\alpha)}(x,y)\chi_{-}(y) &\equiv 0
\end{align*}
as is easily seen by looking at the supports
of $\chi_{+}$, $\chi_{-}$, $\chi_{1}$, and $\chi_{3}$.
Next, we note that the following three terms fulfill the
conditions in Lemma~\ref{lem:decay} and therefore are
\(o(\alpha^{n}),\alpha\to0\), for all \(n\in\mathbb N\):
\begin{align*}
  \chi_{+}(x)L_{1}^{(\alpha)}(x,y)\chi_{-}(y)&\neq0\ \text{ for }
  |x|>2l \text{ and } |y|<(1+\beta)l \\
  \chi_{+}(x)L_{3}^{(\alpha)}(x,y)\chi_{-}(y) &\neq0\ \text{ for } 
  |x|>(1-\beta)\alpha^{t}
  \text{ and } |y|<2l \\
  \chi_{+}(x)L_{2}^{(\alpha)}(x,y)\chi_{-}(y)&\neq0 
  \ \text{ for }|x|>(1-\beta)\alpha^{t} 
  \text{ and } |y|<2l \\
  &\!\!\!\!\text{ and }\ \text{ for }
  |x|\in [2l,(1-\beta)\alpha^{t}]\text{ and }
  |y|<(1+\beta)l .
\end{align*}
This is due to the fact that for $\alpha$ small enough,
$\alpha^{t}>\alpha^{r}$, since $t<2/3<8/9<r$. 
The above is symmetric in $x$ and $y$, which
gives another three terms. 

We are then left with four terms. For these we will 
use that, by the Mean Value Theorem,
$(\chi_{j}(x)-\chi_{j}(y))^{2}
  \leq \|\nabla\chi_{j}\|_{\infty}^{2}|x-y|^{2}$. Note that for the four
remaining terms, 
\begin{align}
 \label{eq:last_terms}
 \chi_{+}L_{2}^{(\alpha)}\chi_{+}\quad ,\quad
 \chi_{-}L_{1}^{(\alpha)}\chi_{-}\quad , \quad
 \chi_{+}L_{3}^{(\alpha)}\chi_{+}\quad , \quad
 \chi_{-}L_{2}^{(\alpha)}\chi_{-}\quad,
\end{align}
we only need to take the supremum of $|\nabla\chi_{j}(\xi)|$ 
over the $\xi$'s between $x$
and $y$ in the support of the relevant term. In this way we get:
\begin{align*}
  |(f,\chi_{\pm}L_{j}^{(\alpha)}\chi_{\pm}f)|
  &\leq 
  \iint |f(x)|\,\chi_{\pm}(x)|f(y)|\,\chi_{\pm}(y) 
  L_{j}^{(\alpha)}(x,y)\,d^{3}\!x\,d^{3}\!y \\
  &\leq
  \frac{c_{j}^{\pm}(\alpha)\alpha^{-2}}{4\pi^{2}}
  \int |f(x)|\,\chi_{\pm}(x)\big(
  (|f|\,\chi_{\pm})*G_{\alpha}\big)(x)\,d^{3}\!x
\end{align*}
where $G_{\alpha}(x)=K_{2}(\alpha^{-1}|x|)$ and 
$c_{j}^{\pm}(\alpha)=\sup_{|x|\gtrless2l}
|\nabla\chi_{j}(x)|^{2}$.
By first using the Cauchy-Schwartz inequality, then Young's
inequality,
we get
\begin{align*}
|(f,\chi_{\pm}L_{j}^{(\alpha)}\chi_{\pm}f)|
  &\leq
  \frac{c_{j}^{\pm}(\alpha)\alpha^{-2}}{4\pi^{2}}
  \|f\,\chi_{\pm}\|_{2}\,\|\,(|f|\,\chi_{\pm})*G_{\alpha}\|_{2}
  \leq
  \frac{c_{j}^{\pm}(\alpha)}{4\pi^{2}\alpha^2}
  \|f\,\chi_{\pm}\|_{2}^{2}\,\|G_{\alpha}\|_{1}.
 \end{align*}
Since
\begin{equation*}
  \|G_{\alpha}\|_{1}=\int K_{2}(\alpha^{-1}|x|)\,d^{3}\!x
  = 4\pi\int_{0}^{\infty}\alpha^{2}t^{2}K_{2}(t)\alpha\,dt
  =6\pi^{2}\alpha^{3}
\end{equation*}
(see (\ref{eq:K_{2}-int}) in Appendix~\ref{appendixA} for 
$\int_{0}^{\infty}t^{2}K_{2}(t)\,dt$) 
we get the following inequality:
\begin{equation}
  \label{eq:temp1}
  |(f,\chi_{\pm}L_{j}^{(\alpha)}\chi_{\pm}f)|
  \leq \frac{3c_{j}^{\pm}(\alpha)\alpha}{2}\|f\,\chi_{\pm}\|_{2}^{2}.
\end{equation}
For two of the terms in \eqref{eq:last_terms}, $\chi_{+}L_{2}^{(\alpha)}\chi_{+}$ and 
$\chi_{+}L_{3}^{(\alpha)}\chi_{+}$, this is sufficient, since
(see \eqref{eq:chis}; recall that \(l=\alpha^{r}\))
\begin{equation*}
  c_{j}^{+}(\alpha)=\sup_{|x|>2l}|\nabla \chi_{j}|^{2} 
= c_{j}^{+}\alpha^{-2t}\quad ,
  \quad j=2,3,
\end{equation*}
and since $t<2/3$ we get, using \eqref{eq:temp1}, that
\begin{equation*}
 \sum_{i=1}^{N}(m_{i},\chi_{+}L_{3}^{(\alpha)}\chi_{+}m_{i})
  \leq N\frac{3}{2}c_{j}^{+}\alpha^{1-2t} = o(\alpha^{-4/3})
  \quad,\quad\alpha\to0,
\end{equation*}
as $N=\lambda Z=\lambda\delta\alpha^{-1}$ ($\lambda$ and $\delta$
fixed) and $\|m_{i}\|_{2}=1$. 
Similarly for
$\chi_{+}L_{2}^{(\alpha)}\chi_{+}$. 

For the other two terms in \eqref{eq:last_terms}, note that
\begin{align*}
  \|f\,\chi_{-}\|_{2}^{2} &=\int |f(x)|^{2}\,|\chi_{-}(x)|^{2}\,d^{3}\!x
  =\int |f(x)|^{2}\,\chi_{-}(x)\,d^{3}\!x 
  =(f,\chi_{-}f)\\
  & = (f,\chi_{-}(\chi_{1}^{2}+\chi_{2}^{2})f)
  =(\chi_{1}f,\chi_{-}\chi_{1}f)+(\chi_{2}f,\chi_{-}\chi_{2}f)\ ,
\end{align*}
since $\chi_{-}^{2}=\chi_{-}$ and $\chi_{1}^{2}+\chi_{2}^{2}=1$
on $\supp\chi_{-}$. Using this and \eqref{eq:temp1}, we obtain (since 
$\chi_{-}\chi_{1}=\chi_{1}$):
\begin{align}
\label{eq:rest_error}
 \sum_{i=1}^{N}(m_{i},\chi_{-}(L_{1}^{(\alpha)}&+L_{2}^{(\alpha)})
 \chi_{-}m_{i})\\
  & \leq 
  C\,\alpha^{1-2r}
 \Big(\sum_{i=1}^{N}(\chi_{1}m_{i},\chi_{1}m_{i})+ 
      \sum_{i=1}^{N}(\chi_{2}m_{i},\chi_{-}\chi_{2}m_{i})\Big)
  \nonumber
\end{align}
where
\begin{align*}
  C&=\frac{3}{2}(c_{1}+c_{2})\quad , \quad
  c_{j}\alpha^{-2r}= \sup_{|x|<2l}|\nabla \chi_{j}(x)|^{2}\ , \ j=1,2.
\end{align*}
The two terms in~(\ref{eq:rest_error}) will be estimated in the 
following two sections, the
first one along with the energy at the nucleus, the second one
with the energy in the intermediary zone.

\section{The energy near the nucleus}
\label{inner}

In this section we estimate the energy at the nucleus, that is (see
\eqref{eq:slater}),
the term
\begin{equation}
 \label{eq:inner}
 \sum_{i=1}^{N}(m_{i},\chi_{1}\big\{T(p)-V(x)
 +\alpha\,\rho*\phi_{a}*|x|^{-1}\big\}\chi_{1}m_{i}).
\end{equation}
Also, half of the remaining term~(\ref{eq:rest_error})
of the localisation error, 
treated in the previous section, will be estimated here.
We start by noting that $\rho*\phi_{a}*|x|^{-1}$ is positive, so that we get
a lower bound to~(\ref{eq:inner}) by dropping this term.
The remaining expression will be treated by using the following 
result by Lieb and Yau~\cite[Theorem 11]{LiYau88} on the
hyper-relativistic operator $|p|$:
\begin{thm}
 \label{thm:LiYau}
 Let $C_{0}>0$ and $R>0$ and let
\begin{equation*}
  H_{C_{0}R}=|p|-\frac{2}{\pi}|x|^{-1}-C_{0}/R
\end{equation*}
be defined on $L^2(\R^3)$ as a quadratic form. Let $0\leq\gamma\leq q$
be a density matrix (that is, any bounded operator on  $L^2(\R^3)$
which satisfies the operator inequality $0\leq\gamma\leq q$ and
for which $\Tr(\gamma)<\infty$) and let $\chi$ 
be any function with support in 
$B_{R}=\{x\, |\,|x|\leq R\}$. Then
\begin{equation}
\label{eq:LiYau}
  \Tr(\bar\chi\gamma\chi H_{C_{0}R}) \geq 
  - 4.4827\,C_{0}^4R^{-1}q\{(3/4\pi R^3)\int |\chi(x)|^2\,d^{3}\!x\}.
\end{equation}
Note, that when $\chi\equiv1$ in $B_{R}$, then the factor in braces
\{\} in \eqref{eq:LiYau} is 1.
\end{thm}
Here, $\Tr(\gamma h)$ is shorthand for $\sum_{k}(f_{k},hf_{k})\gamma_{k}$,
where $(f_{k},\gamma_{k})$ are the eigenfunctions and eigenvalues of
$\gamma$. For more details, see Lieb
\cite{Li83}. In our situation, $q=2$.
For our purpose, let $\Pi$ be the projection on span$\{m_{i}\,|\,i=1,\dots
N\}$, then $\Pi$ is a density matrix as above, and
\begin{equation*}
  \Tr(\chi_{1}\Pi\chi_{1} H_{C_{0}R}) = \sum_{i=1}^{N} (m_{i},
  \chi_{1} H_{C_{0}R}\chi_{1}
  m_{i}).
\end{equation*}
Since $\supp\chi_{1}\subseteq B(0,(1+\beta)\alpha^{r})$ 
with $8/9<r<1$, set
$R=(1+\beta)\alpha^{r}$ and $C_{0}=2(1+\beta)\alpha^{r-1}$. Then
\begin{align*}
  T(p)-V(x) &=\sqrt{p^{2}+\alpha^{-2}}-\alpha^{-1}-\frac{\delta}{|x|}\\
  &\geq
  |p|-\alpha^{-1}-\frac{2}{\pi}|x|^{-1}=H_{C_{0}R}+\alpha^{-1},
\end{align*}
since $\sqrt{p^{2}+\alpha^{-2}}-\alpha^{-1}\geq|p|-\alpha^{-1}$ and
$\delta\leq2/\pi$.
Including the first term in~(\ref{eq:rest_error}) we
now have, applying \eqref{eq:LiYau},
\begin{align}
  \sum_{i=1}^{N}(m_{i},\chi_{1}&
  \big\{T(p)-V(x)\big\}\chi_{1}m_{i}) - 
  C\,\alpha^{1-2r}
  \sum_{i=1}^{N}(m_{i},\chi_{1}\chi_{1}m_{i}) \nonumber \\
  &\geq 
  \sum_{i=1}^{N}(m_{i},\chi_{1}\big\{H_{C_{0}R}+\alpha^{-1}-
  C\,\alpha^{1-2r}\big\}
  \chi_{1}m_{i})\nonumber \\
  &\geq 
  \sum_{i=1}^{N}(m_{i},\chi_{1}H_{C_{0}R}
  \chi_{1}m_{i})  
  = \Tr(\bar\chi_{1}\Pi\chi_{1}H_{C_{0}R})\geq
  -C\alpha^{3r-4}.
  \label{eq:rest1}
\end{align}
The second inequality is valid for $\alpha$ small enough,
since $r<1$, so that $\alpha^{2(1-r)}\to 0$ for $\alpha\to 0$.
 Since $3r-4>-4/3$ (as \(8/9<r\)),  
the RHS of~\eqref{eq:rest1} is $o(\alpha^{-4/3})$, $\alpha\to 0$,
which is the desired order.
Note that the above procedure is
what decides the scale $\alpha^{r}$, $8/9<r<1$, 
on which one can localise near the nucleus.

\section{The intermediary zone}
\label{inter}

The energy in this area is given by the term (see \eqref{eq:slater})
\begin{equation}
 \label{eq:inter}
 \sum_{i=1}^{N}(m_{i},\chi_{2}\big\{T(p)-V(x)
 +\alpha\,\rho*\phi_{a}*|x_{i}|^{-1}\big\}\chi_{2}m_{i}).
\end{equation}
The zone defined by the $\chi_{2}$ was introduced to separate
the outer zone defined by $\chi_{3}$ and the support of the 
coherent states to be 
introduced later. As in the previous section we note that
by dropping the term involving $\rho*\phi_{a}*|x|^{-1}$, we get a
lower bound of the energy in~(\ref{eq:inter}). 
The remaining expression will be estimated 
by a generalisation
of the Lieb-Thirring inequality (see Lieb and
Thirring~\cite{LiThi75}), 
proved by Daubechies 
in~\cite[page 518]{Dau83}. See also page 516 loc.\ cit.\ 
for the conditions on the function $T(p)$.
\begin{prop}
Let $F(s)=\int_{0}^{s}dt\,[T^{-1}(t)]^{3}$, where
$T(p)=T(|p|)=\sqrt{|p|^{2}+\alpha^{-2}}-\alpha^{-1}$ as a function. Then
\begin{equation*}
  \langle\psi,\sum_{i=1}^{N}\big\{ T(p_{i})-V(x_{i})\big\}\psi\rangle
  \geq{}-q\tilde C\!\!\int F(|V(x)|)\,d^{3}\!x
\end{equation*}
where $\tilde C\leq 0.163$.
\end{prop}
Note that this in particular means that the negative part
of the spectrum of the operator $T(p)-V(x)$ is discrete and that the
sum of the negative eigenvalues of this operator 
is bounded from below by the quantity
$-q\tilde C\!\int F(|V(x)|)\,d^{3}\!x$. To see this, let 
$\{e_{j}\}_{j=0}^{\infty}$ be
these negative eigenvalues, $e_{0}\leq e_{1}\leq\dots$,
and $\{g_{j}\}_{j=0}^{\infty}$ corresponding orthonormal
eigenfunctions, 
and let
$\psi$ be the Slater-determinant of the first $N$ of the
$g_{j}$'s. Then, by the above proposition, 
\begin{align}
\label{eq:est_eigenvalues}
  {}-q\tilde C\!\!\int F(|V(x)|)\,d^{3}\!x\,&\leq
  \langle\psi,\sum_{i=1}^{N}\big\{ T(p_{i})-V(x_{i})\big\}
  \psi\rangle\notag\\
  &= \sum_{j=1}^{N}(g_{j}, \big\{T(p)-V(x)\big\}g_{j}) = \sum_{j=1}^{N}e_{j}.
\end{align}
Since the left-hand-side is independent of $N$, we get the statement
by taking the limit $N\to\infty$.
This will, as mentioned above, be used on the energy related to the
cut-off $\chi_{2}$, but also on the remaining half of the term 
$\chi_{-}(L_{1}^{(\alpha)}+L_{2}^{(\alpha)})\chi_{-}$ 
discussed in Section~\ref{loc_error}, see~\eqref{eq:rest_error}. First,
let us calculate $F$:
\begin{equation*}
  T(p)=T(|p|)=\sqrt{|p|^{2}+\alpha^{-2}}-\alpha^{-1}
  \Rightarrow T^{-1}(t)=\sqrt{t^{2}+2\alpha^{-1}t}.
\end{equation*}
Then
\begin{align*}
F(s)=\int_{0}^{s}(t^{2}+2\alpha^{-1}t)^{3/2}\,dt
    =\int_{0}^{s}\big(\frac{2t}{\alpha}\big)^{3/2}
    \big(1+\frac{\alpha t}{2}\big)^{3/2}\, dt.
\end{align*}
Now, by a Taylor expansion of the second term in the integral, we get
\begin{align}
\label{eq:Taylor}
  \big(1+(\alpha t)/2\big)^{3/2}
   \leq
  1+\frac{3\alpha}{4}t+\frac{3\alpha^{2}}{32}t^{2}.
\end{align}
That is, for $s\geq0\,$:
\begin{align}
  \label{eq:est_F}
  F(s)\leq \big(\frac{2}{\alpha}\big)^{3/2}
  \Big\{\frac{2}{5}s^{5/2}+\frac{3\alpha}{14}s^{7/2}
  +\frac{\alpha^{2}}{48}s^{9/2}\Big\}.
\end{align}
The two terms we wish to estimate in this section are, as mentioned
above, 
\begin{align*}
  \sum_{i=1}^{N}(m_{i},\chi_{2}\big\{T(p)-V(x)\big\}\chi_{2}m_{i})
  \quad\text{ and }\quad
   C\alpha^{1-2r}
  \sum_{i=1}^{N}(\chi_{2}m_{i},\chi_{-}\chi_{2}m_{i}).
\end{align*}
In order to do so, note that on $\supp\chi_{-}\chi_{2}$ we have
(\(\chi_{-}\) being the characteristic function of \(B(0,2\alpha^{r})\))
\begin{equation*}
 V(x) = \frac{\delta}{|x|}\geq \frac{\delta}{2\alpha^{r}}
 \geq C\,\alpha^{1-2r}
\end{equation*}
for $\alpha$ small enough, since $r<1$, so that $\alpha^{1-r}\to 0$
as $\alpha\to 0$. 
Therefore, by the estimate~\eqref{eq:est_F} on $F(s)$, and still for $\alpha$
small enough, we have
\begin{align}
  \label{est:hatV}
  \sum_{i=1}^{N}(m_{i},\chi_{2}\big\{T(p)&-V(x)\big\}\chi_{2}m_{i})
  - C\,\alpha^{1-2r} \sum_{i=1}^{N}(m_{i},
  \chi_{2}\chi_{-}\chi_{2}m_{i})\nonumber\\
  &\geq 
  \sum_{i=1}^{N}(m_{i},\chi_{2}\big\{ 
  T(p)-2\hat V(x)\big\} \chi_{2}m_{i})
\end{align}
with $\hat V(x)=\chi_{2}(x)V(x)$. Letting $(e_{j},g_{j})$ be the
negative eigenvalues and corresponding orthonormal eigenvectors 
for the operator $T(p)-2\hat V(x)$ as before, we then have
\begin{align}
 \label{est:hatVbis}
 \sum_{i=1}^{N}&(m_{i},\chi_{2}\big\{ 
  T(p)-2\hat V(x)\big\} \chi_{2}m_{i})
  \geq \sum_{i=1}^{N}(\chi_{2}m_{i},\Big\{\sum_{j}e_{j}(g_{j},\,\cdot \,
  )\,g_{j}\Big\}\,\chi_{2}m_{i})\\
  &=\sum_{j}\sum_{i=1}^{N}e_{j}
  |(\chi_{2}m_{i},g_{j})|^{2}=\sum_{j}\sum_{i=1}^{N}e_{j}
  |(m_{i},\chi_{2}g_{j})|^{2}
  \geq \sum_{j}e_{j}\|\chi_{2}g_{j}\|^{2}
  \geq \sum_{j}e_{j}.\nonumber
\end{align}
Here we used Bessel's inequality (remember that the $m_{i}$'s are
orthonormal), that $e_{j}<0$ and that
$0\leq\chi_{2}\leq1$. Using~\eqref{eq:est_eigenvalues} on $T(p)-2\hat
V(x)$, in
the limit $N\to\infty$, we now reach (using \eqref{eq:est_F},
\eqref{est:hatV}, and \eqref{est:hatVbis})
\begin{align*}
  \sum_{i=1}^{N}&(m_{i},\chi_{2}\big\{T(p)-V(x)\big\}\chi_{2}m_{i})
  - C\,\alpha^{1-2r} \sum_{i=1}^{N}(m_{i},
  \chi_{2}\chi_{-}\chi_{2}m_{i})\\
  &\geq
  {}-2\tilde C \!\!\!\!\!\!\int\limits_
  {\supp\ \chi_{2}}\!\!\!\!F(2\,|V(x)|)\,d^{3}\!x\\\
  &\geq
  {}-C \!\!\!\!\!\!\int\limits_{\supp\ \chi_{2}}\!\!\!\!\!
  \Big(\frac{2}{\alpha}\Big)^{3/2}\!
  \Big\{\frac{2}{5}\big(2|V(x)|\big)^{5/2}\!\!+\!\frac{3\alpha}{14}
  \big(2|V(x)|\big)^{7/2}
  \!\!
  +\frac{\alpha^{2}}{48}
  \big(2|V(x)|\big)^{9/2}\Big\}\,d^{3}\!x\\
  &= -C 4\pi \int_{\alpha^{r}}^{\alpha^{t}}
  \Big(\frac{2}{\alpha}\Big)^{3/2}
  \Big\{\frac{2}{5}\big(\frac{2\delta}{|x|}\big)^{5/2}+\frac{3\alpha}{14}
  \big(\frac{2\delta}{|x|}\big)^{7/2}
  +\frac{\alpha^{2}}{48}
  \big(\frac{2\delta}{|x|}\big)^{9/2}\Big\}
  |x|^{2}\,d|x|\\
  &={} -C\Big[\,\frac{4}{5}(\alpha^{\frac{t-3}{2}}-\alpha^{\frac{
  r-3}{2}})+\frac{6\delta}{7}(\alpha^{\frac{-(r+1)}{2}}-\alpha^{\frac{
  -(t+1)}{2}})\
  +\frac{4\delta^{2}}{72}(\alpha^{\frac{1-3r}{2}}
  -\alpha^{\frac{
  1-3t}{2}})\Big].
\end{align*}
Since $8/9<r<1$ and
$1/3<t<2/3$ , all of these terms are $o(\alpha^{-4/3})$, which is
the desired order. We note that it is this analysis that decides
the scale $\alpha^{t}$ of the outer cut-off $\chi_{3}$.

\section{The outer zone and Thomas-Fermi teory}
\label{outer}

Up to order $o(\alpha^{-4/3})$ we are now left with 
\begin{align*}
  \sum_{i=1}^{N}(m_{i}&,\chi_{3}\big\{T(p)-V(x)
  +\alpha\,\rho*\phi_{a}*|x|^{-1}\big\}\chi_{3}m_{i})\\
  &-\frac{\alpha}{2}\iint\frac{\rho(x)\rho(y)}{|x-y|}\,d^{3}\!xd^{3}\!y
  -\alpha\,c(\phi)Na^{-1}.
\end{align*}
This expression will now be related to the semi-classical Thomas-Fermi
energy. This is done by introducing coherent states, following
Lieb and Yau in~\cite[proof of Lemma B.3]{LiYau87}. Let $g$ 
be the function chosen in Section~\ref{one_par}, that is, $g\in
C_{0}^{\infty}(\R^{3})$, spherically symmetric,  non-negative, 
supported in the unit ball
$B(0,1)$ of $\R^{3}$ and such that
$\int\!g(x)^{2}\,d^{3}\!x=1$. Let 
$g_{\alpha}(x)=\alpha^{-3s/2}g(x/\alpha^{s})$, $1/3<t<s<2/3$, that is,
$\phi_{a}(x)=g_{\alpha}(x)^{2}$ with $a=\alpha^{s}$. In this way, since 
$N=\lambda Z=\lambda\delta\alpha^{-1}$:
\begin{equation*}
  \alpha\,c(\phi)Na^{-1}=\lambda\delta\,c(\phi)\alpha^{-s}=o(\alpha^{-2/3}), 
\end{equation*}
which is also $o(\alpha^{-4/3}), \alpha\to0$.
Define now the coherent
states $g_{\alpha}^{p,q}$, $p,q\in\R^{3}$ by
\begin{equation*}
  g_{\alpha}^{p,q}(x)=g_{\alpha}(x-q)e^{ipx}.
\end{equation*}
With $\tilde T(p)$ the function $\sqrt{p^{2}+\alpha^{-2}}-\alpha^{-1}$, 
we then have the formulae
\begin{align}
\label{eq:coh_states}
  (f,f)&=\frac{1}{(2\pi)^{3}}\iint
  d^{3}\!p\,d^{3}\!q\,(f,g_{\alpha}^{p,q})
  (g_{\alpha}^{p,q},f), \notag\\
  (f, (V*g_{\alpha}^{2})f) &= 
  \frac{1}{(2\pi)^{3}}\iint d^{3}\!p\,d^{3}\!q\,V(q)(f,g_{\alpha}^{p,q})
  (g_{\alpha}^{p,q},f),\\
  (f, T(p)f) &\geq 
  \frac{1}{(2\pi)^{3}}\iint d^{3}\!p\,d^{3}\!q\,\tilde T(p)
  (f,g_{\alpha}^{p,q})
  (g_{\alpha}^{p,q},f)
  -o(\alpha^{-1/3}).\notag
\end{align}
The proof of these formulae is carried out in 
Appendix~\ref{appendixB}.
In this way, letting $\tilde V(x)=\delta/|x| - \alpha\,\rho*|x|^{-1}$
(remember that $\phi_{\alpha^{s}}=g_{\alpha}^{2}$):
\begin{align*}
 \sum_{i=1}^{N}&(m_{i},\chi_{3}\big\{T(p)-V(x)+
 \alpha\,\rho*\phi_{a}*|x|^{-1}\big\}\chi_{3}m_{i}) \\
 &= \sum_{i=1}^{N}(m_{i},\chi_{3}\big\{T(p)-\tilde
 V(x)*\phi_{\alpha^{s}}+\frac{\delta}{|x|}*\phi_{\alpha^{s}}
 -\frac{\delta}{|x|}\big\}\chi_{3}m_{i}) \\
 &=\sum_{i=1}^{N}(m_{i},\chi_{3}\big\{T(p)-\tilde
 V(x)*\phi_{\alpha^{s}}\big\}\chi_{3}m_{i}) \\
 &=\frac{1}{(2\pi)^{3}}\iint d^{3}\!p\,d^{3}\!q\,
 \big(\tilde T(p)-\tilde V(q)\big)\big(
 \sum_{i=1}^{N}|(m_{i}\chi_{3},g_{\alpha}^{p,q})|^2\big) 
 -N\,o(\alpha^{-1/3}).
\end{align*}
The second equality follows from Newton's theorem (since
$\phi_{\alpha^{s}}$
is spherically symmetric):
$|x|^{-1}-|x|^{-1}*\phi_{\alpha^{s}}\equiv0$ outside 
$\supp\ \phi_{\alpha^{s}}$, and since 
$\supp\ \chi_{3}\cap\supp\ \phi_{\alpha^{s}}=\emptyset$ for 
$\alpha$ sufficiently small (as $s>t$), 
\begin{equation*}
 \sum_{i=1}^{N}(m_{i},\chi_{3}\big\{\frac{\delta}{|x_{i}|}*\phi_{\alpha^{s}}
 -\frac{\delta}{|x_{i}|}\big\}\chi_{3}m_{i}) = 0.
\end{equation*}
This is one of the reasons for introducing the
intermediary
zone by the function $\chi_{2}$. Note also that 
$N\,o(\alpha^{-1/3}) = o(\alpha^{-4/3})$.
Now, for $\alpha$ small enough, $\alpha^{s-t}<1/4$, since 
$s>t$, so that if $|q|<\frac{1}{4}\alpha^{t}$, then
\begin{equation*}
  |x-q|<\alpha^{s} \Rightarrow |x|<\frac{1}{2}\alpha^{t},
\end{equation*}
and so $(m_{i}\chi_{3},g_{\alpha}^{p,q})=0$, since $\supp\ g_{\alpha}\subset
B(0,\alpha^{s})$ and 
$\supp\ \chi_{3}\subset\R^{3}\setminus B(0,\frac{1}{2}\alpha^{t})$.
That is, for $\alpha$ small enough
\begin{equation*}
 \supp_{q} |(m_{i}\chi_{3},g_{\alpha}^{p,q})|^{2} \subseteq \R^{3}\setminus
 B(0,\tfrac{1}{4}\alpha^{t}),
 \end{equation*}
so that for any $\mu\geq0$ we have, with
$M(p,q)=\sum_{i=1}^{N}|(m_{i}\chi_{3},g_{\alpha}^{p,q})|^2$ and
$\big[f\big]_{\pm}=\max\{\pm f,0\}$:
\begin{align*}
  \frac{1}{(2\pi)^{3}}&\iint d^{3}\!p\,d^{3}\!q\,\big(\tilde T(p)-
  \tilde V(q)\big)\big(
  \sum_{i=1}^{N}|(m_{i}\chi_{3},g_{\alpha}^{p,q})|^2\big)\\
  &=\frac{1}{(2\pi)^{3}}\iint\limits_{|q|>\frac{1}{4}\alpha^{t}}
  d^{3}\!p\,d^{3}\!q\,
  \big(\tilde T(p)-(\tilde V(q)-\alpha\mu)\big) \,M(p,q) 
  -\alpha\mu\sum_{i=1}^{N}(\chi_{3}m_{i},\chi_{3}m_{i})\\
  &\geq -\frac{1}{(2\pi)^{3}}\iint\limits_{|q|>\frac{1}{4}\alpha^{t}}
  d^{3}\!p\,d^{3}\!q\,
  \Big[ \tilde T(p)-(\tilde V(q)-\alpha\mu)\Big]_{-} -\alpha\mu N,
\end{align*}
since $0\leq M(p,q)\leq1$ and 
$(\chi_{3}m_{i},\chi_{3}m_{i})\leq\|m_{i}\|_{2}^{2}=1$. 
The first is seen by Bessel's inequality, since
the $m_{i}$'s are orthonormal and 
$\|\chi_{3}g_{\alpha}^{p,q}\|_{2}\leq\|g_{\alpha}^{p,q}\|_{2}=1$.
In this way we have shown that for $\mu\geq0$, $\rho:\R^{3}\to\R$ and
$\psi\in{\mathcal H}_F=\bigwedge^N L^2(\R^3,\C^2)\,$:
\begin{align}
\label{eq:prelimi}
  \langle\psi,H\psi\rangle&\geq
  -\frac{1}{(2\pi)^{3}}\iint\limits_{|q|>\frac{1}{4}\alpha^{t}}
  d^{3}\!p\,d^{3}\!q\,
  \Big[\tilde T(p)-(\tilde V(q)-\alpha\mu)\Big]_{-} \notag\\
  &\qquad-\frac{\alpha}{2}\iint
  \frac{\rho(x)\rho(y)}{|x-y|}\,d^{3}\!x\,d^{3}\!y
  -\alpha\mu N -o(\alpha^{-4/3}).
\end{align} 
Choose now $\rho$ to be the Thomas-Fermi density $\rho_{TF}^{N,Z}$, that
is, the function that minimises the Thomas-Fermi functional (here, 
$\gamma=(3\pi^{2})^{2/3}$):
\begin{equation}
\label{eq:TF-func}
  \mathcal{E}_{TF}(\rho)=\frac{3}{5}\gamma\int\rho(x)^{5/3}\,d^{3}\!x
  -\int\rho(x)\frac{Z}{|x|}\,d^{3}\!x
  +\frac{1}{2}\iint\frac{\rho(x)\rho(y)}{|x-y|}\,d^{3}\!x\,d^{3}\!y
\end{equation}
over the set
\begin{equation*}
\Big\{\rho\in L^{5/3}(\R^{3})\cap L^{1}(\R^{3})\,
  \big|\,\rho\geq0,\int\rho(x)\,d^{3}\!x\leq N\Big\}.
\end{equation*}
(For the Thomas-Fermi theory, see Lieb and Simon~\cite{LiSi77} and
Lieb~\cite{Li81b}).
Then $\rho_{TF}^{N,Z}$ satisfies the Thomas-Fermi equation:
\begin{equation}
\label{eq:TF-eq}
  \gamma\,\rho(x)^{2/3}=\Big[\frac{Z}{|x|}-\rho*|x|^{-1}-\mu\Big]_{+}
\end{equation}
for some unique $\mu=\mu(N)$. Furthermore,  
\begin{align*}
  &\text{for $N\leq Z$: }\quad \int\rho_{TF}^{N,Z}(x)\,d^{3}\!x=N
  \quad\text{and}\quad\mu(N)>0,\\ 
  &\text{for $N>Z$: }\quad 
  \int\rho_{TF}^{N,Z}(x)\,d^{3}\!x=Z \quad\text{and}\quad\mu(N)=0
\end{align*}
(see Lieb and Simon~\cite[Theorems II.17, 18 and 20]{LiSi77}).
In this way, $\int\rho_{TF}^{N,Z}(x)\,d^{3}\!x<N$ implies $N>Z$, 
and therefore $\mu(N)=0$, so that we always have
\begin{equation}
\label{eq:mu-rho-N}
  \mu(N)\int\rho_{TF}^{N,Z}(x)\,d^{3}\!x=\mu(N)N.
\end{equation}
Let $\mathcal{E}_{TF}(N,Z)\equiv\mathcal{E}_{TF}(\rho_{TF}^{N,Z})$ 
and define the Thomas-Fermi potential by
\begin{equation*}
  V_{TF}^{N,Z}(x)\equiv Z/|x|-\rho_{TF}^{N,Z}*|x|^{-1}-\mu(N),
\end{equation*}
then we have the following scaling (\cite[(2.24) p.608]{LiSi77})
(remember, that $\lambda=N/Z$ is fixed):
\begin{align}
\label{eq:scale_E}
 \mathcal{E}_{TF}(N,Z)&=Z^{7/3}\mathcal{E}_{TF}(\lambda,1) 
 \equiv -C_{TF}(\lambda)Z^{7/3},\\
 \label{eq:scale_V}
 V_{TF}^{N,Z}(x)&=Z^{4/3}V_{TF}^{\lambda,1}(Z^{1/3}x)\equiv
 Z^{4/3}V_{TF}(Z^{1/3}x).
\end{align}

The idea is now to estimate the difference between the integral
in~(\ref{eq:prelimi}) (with $\rho=\rho_{TF}^{N,Z}$ and $\mu=\mu(N)$)
and
\begin{equation*}
  -\frac{\alpha}{(2\pi)^{3}}\iint\limits_{|q|>\frac{1}{4}\alpha^{t}}
  d^{3}\!p\,d^{3}\!q\,
  \Big[ \frac{p^{2}}{2}-\big(\frac{Z}{|q|}-
  \rho_{TF}^{N,Z}*|q|^{-1}-\mu(N)\big)\Big]_{-} .
\end{equation*}
This is done in two steps: first, we change the domain of the
integration,
then we change the integrand, each time estimating the error.

First,
\begin{eqnarray*}
  -\frac{1}{(2\pi)^{3}}\iint\limits_{|q|>\frac{1}{4}\alpha^{t}}
  d^{3}\!p\,d^{3}\!q\,
  \Big[ \tilde T(p)-\alpha V_{TF}^{N,Z}(q)\Big]_{-} 
  =\frac{1}{(2\pi)^{3}}\!\!\!\!\!\!\!\!\!\!\!\!\!\!\!\!\!\!
  \iint\limits_{|q|>\frac{1}{4}\alpha^{t};\ 
  \tilde T(p)<\alpha V_{TF}^{N,Z}(q)}
  \!\!\!\!\!\!\!\!\!\!\!\!\!\!\!\!\!
  d^{3}\!p\,d^{3}\!q\,
  \big(\tilde T(p)-\alpha V_{TF}^{N,Z}(q)\big)\\
  =\frac{1}{(2\pi)^{3}}\!\!\!\!\!\!\!\!\!\!\!\!\!\!\!\!\!
  \iint\limits_{|q|>\frac{1}{4}\alpha^{t};\ 
  \alpha\frac{p^{2}}{2}<\alpha V_{TF}^{N,Z}(q)}
  \!\!\!\!\!\!\!\!\!\!\!\!\!\!\!\!\!
  d^{3}\!p\,d^{3}\!q\,
  \big(\tilde T(p)-\alpha V_{TF}^{N,Z}(q)\big) + 
  \frac{1}{(2\pi)^{3}}
  \!\!\!\!\!\!\!\!\!\!\!\!\!\!\!\!\!\!\!\!\!\!\!
  \iint\limits_{|q|>\frac{1}{4}\alpha^{t};\
  \tilde T(p)<\alpha V_{TF}^{N,Z}(q)<\alpha\frac{p^{2}}{2}}
  \!\!\!\!\!\!\!\!\!\!\!\!\!\!\!\!\!\!\!\!\!\!\!
  d^{3}\!p\,d^{3}\!q\,
  \big(\tilde T(p)-\alpha V_{TF}^{N,Z}(q)\big).
\end{eqnarray*}
Since $\tilde T(p)\geq0$, we get
\begin{align*}
  \iint\limits_{|q|>\frac{1}{4}\alpha^{t};\
  \tilde T(p)<\alpha V_{TF}^{N,Z}(q)<\alpha\frac{p^{2}}{2}}
  \!\!\!\!\!\!\!\!\!\!\!\!\!\!\!\!\!\!\!\!\!\!\!
  d^{3}\!p\,d^{3}\!q\,
  \big(\alpha V_{TF}^{N,Z}(q)&-\tilde T(p)\big)
  \leq
  \alpha\!\!\!\!\!\!\!\!\!\!\!\!\!\!\!\!\!\!\!\!\!\!\!
  \iint\limits_{|q|>\frac{1}{4}\alpha^{t};\
  \tilde T(p)<\alpha V_{TF}^{N,Z}(q)<\alpha\frac{p^{2}}{2}}
  \!\!\!\!\!\!\!\!\!\!\!\!\!\!\!\!\!\!\!\!\!\!\!
  d^{3}\!p\,d^{3}\!q\,
  V_{TF}^{N,Z}(q).
\end{align*}
Using the scaling~(\ref{eq:scale_V}) and the change of variables
$\omega=\delta^{1/3}\alpha^{-1/3}q$, the above is equal to
\begin{align}
\label{eq:area_1}
  \delta^{1/3}\alpha^{2/3}
  \!\!\!\!\!\!\!\!\!\!\!\!\!\!\!\!\!\!\!\!\!\!\!
  &\iint\limits_{\substack{|\omega|>\frac{1}{4}\delta^{1/3}\alpha^{t-1/3}\\
  \tilde T(p)<\delta^{4/3}\alpha^{-1/3}V_{TF}(\omega)<\alpha\frac{p^{2}}{2}}}
  \!\!\!\!\!\!\!\!\!\!\!\!\!\!\!\!\!\!\!\!\!\!
  d^{3}\!p\,d^{3}\!\omega\,
  V_{TF}(\omega).
\end{align}
The limits in the integral means that
\begin{equation*}
  2\delta^{4/3}\alpha^{-4/3}V_{TF}(\omega)\leq p^{2}
  \leq  2\delta^{4/3}\alpha^{-4/3}V_{TF}(\omega)
  \big(1+\frac{1}{2}\delta^{4/3}\alpha^{2/3}V_{TF}(\omega)\big)
\end{equation*}
so that with 
\begin{align*}
 X= 2\delta^{4/3}\alpha^{-4/3}V_{TF}(\omega)\ ,\ 
 Y=\frac{1}{2}\delta^{4/3}\alpha^{2/3}V_{TF}(\omega)\ ,\ 
 Z=|p|^{2}\ ,\ 
 W=\frac{1}{4}\delta^{1/3}\alpha^{t-1/3},
\end{align*}
we have
\begin{align*}
 \text{(\ref{eq:area_1})}&=(4\pi)^{2}\delta^{1/3}\alpha^{2/3}
 \int_{W}^{\infty}d|\omega|\,|\omega|^{2}\,V_{TF}(\omega)
 \Big(\int_{X}^{X(1+Y)}\frac{\sqrt{Z}}{2}\,dZ\Big)\\
 &=(4\pi)^{2}\delta^{1/3}\alpha^{2/3}
 \int_{W}^{\infty}d|\omega|\,|\omega|^{2}\,V_{TF}(\omega)
 \frac{X^{3/2}}{3}\big((1+Y)^{3/2}-1\big).
\end{align*}
By the Taylor-expansion~(\ref{eq:Taylor}), we have
 \( (1+Y)^{3/2}\leq1+\frac{3}{2}Y+\frac{3}{8}Y^{2}\),
and so
\begin{align*}
  \text{(\ref{eq:area_1})}&\leq
  C\delta^{7/3}\alpha^{-4/3}
  \int_{W}^{\infty}|\omega|^{2}\,V_{TF}(\omega)^{5/2}\\
  &\qquad\qquad\qquad\qquad\times
  \big(\frac{3}{4}\delta^{4/3}\alpha^{2/3}V_{TF}(\omega)
  +\frac{3}{32}\delta^{8/3}\alpha^{4/3}V_{TF}(\omega)^{2}\big)
  \,d|\omega|.
\end{align*}  
Using that $V_{TF}^{N,Z}(x)\leq Z/|x|$, since $\mu(N)\geq0$ and
$\rho_{TF}^{N,Z}\geq0$ (remember that $V_{TF}\equiv V_{TF}^{\lambda,1}$),
we arrive at
\begin{align*}
  \text{(\ref{eq:area_1})}&\leq
  C\delta^{11/3}\alpha^{-2/3}\int_{W}^{\infty}d|\omega|\,
  |\omega|^{-3/2}+\sqrt{2}\pi^{2}\delta^{5}\int_{W}^{\infty}d|\omega|\,
  |\omega|^{-5/2}\\
  &\sim\alpha^{-2/3}W^{-1/2}+W^{-3/2}\sim
  \alpha^{-2/3}\alpha^{1/6-t/2}+\alpha^{(1-3t)/2}\\
  &= o(\alpha^{-5/6})+ o(\alpha^{-1/2})
\end{align*}
since $t<2/3$.
This means, that
\begin{align*}
  -\frac{1}{(2\pi)^{3}}&\iint\limits_{|q|>\frac{1}{4}\alpha^{t}}
  d^{3}\!p\,d^{3}\!q\,
  \Big[ \tilde T(p)-\alpha V_{TF}^{N,Z}(q)\Big]_{-} \\
  &\geq
  \frac{1}{(2\pi)^{3}}\!\!\!\!\!\!\!\!\!\!\!\!\!\!\!\!\!
  \iint\limits_{|q|>\frac{1}{4}\alpha^{t};\ 
  \alpha\frac{p^{2}}{2}<\alpha V_{TF}^{N,Z}(q)}
  \!\!\!\!\!\!\!\!\!\!\!\!\!\!\!\!\!
  d^{3}\!p\,d^{3}\!q\,
  \big( \tilde T(p)-\alpha V_{TF}^{N,Z}(q)\big) - o(\alpha^{-4/3}).
\end{align*}

Next note that since $|q|>\frac{1}{4}\alpha^{t}$
and $\alpha V_{TF}^{N,Z}(q)\leq\delta/|q|$ 
in the area of integration, we here have that
\begin{equation*}
 \tilde T(p)=\sqrt{p^{2}+\alpha^{-2}} -\alpha^{-1}
 \geq \alpha\frac{p^{2}}{2}-\alpha^{3}\frac{p^{4}}{8}.
\end{equation*}
In this way, we get
\begin{align}
\label{eq:clas_E}
  \frac{1}{(2\pi)^{3}}\!\!\!\!\!\!\!\!\!\!\!\!\!\!\!\!\!\!
  &\iint\limits_{|q|>\frac{1}{4}\alpha^{t};\ 
  \alpha\frac{p^{2}}{2}<\alpha V_{TF}^{N,Z}(q)}
  \!\!\!\!\!\!\!\!\!\!\!\!\!\!\!\!\!\!
  d^{3}\!p\,d^{3}\!q\,
  \big(\tilde T(p)-\alpha V_{TF}^{N,Z}(q)\big) \notag\\
  &\geq\frac{1}{(2\pi)^{3}}\!\!\!\!\!\!\!\!\!\!\!\!\!\!\!\!\!\!
  \iint\limits_{|q|>\frac{1}{4}\alpha^{t};
  \ \alpha\frac{p^{2}}{2}<\alpha V_{TF}^{N,Z}(q)} 
  \!\!\!\!\!\!\!\!\!\!\!\!\!\!\!\!\!\!
  d^{3}\!p\,d^{3}\!q\,
  \big(\alpha\frac{p^{2}}{2}
  -\alpha^{3}\frac{p^{4}}{8}-\alpha V_{TF}^{N,Z}(q)\big)\notag\\
  &=\frac{1}{(2\pi)^{3}}\!\!\!\!\!\!\!\!\!\!\!\!\!\!\!\!\!\!
  \iint\limits_{|q|>\frac{1}{4}\alpha^{t};
  \ \alpha\frac{p^{2}}{2}<\alpha V_{TF}^{N,Z}(q)} 
  \!\!\!\!\!\!\!\!\!\!\!\!\!\!\!\!\!\!
  d^{3}\!p\,d^{3}\!q\,
  \big(\alpha\frac{p^{2}}{2}-\alpha V_{TF}^{N,Z}(q)\big)
  -\alpha^{3}\frac{1}{(2\pi)^{3}}
  \!\!\!\!\!\!\!\!\!\!\!\!\!\!\!\!\!\!
  \iint\limits_{|q|>\frac{1}{4}\alpha^{t};
  \ \alpha\frac{p^{2}}{2}<\alpha V_{TF}^{N,Z}(q)}
  \!\!\!\!\!\!\!\!\!\!\!\!\!\!\!\!
  \frac{p^{4}}{8}\,d^{3}\!p\,d^{3}\!q.
\end{align}
Note that
\begin{align*}
 \frac{1}{(2\pi)^{3}}
  \!\!\!\!\!\!\!\!\!\!\!\!\!\!\!\!\!
  &\iint\limits_{|q|>\frac{1}{4}\alpha^{t};
  \ \alpha\frac{p^{2}}{2}<\alpha V_{TF}^{N,Z}(q)}
  \!\!\!\!\!\!\!\!\!\!\!\!\!\!\!
  d^{3}\!p\,d^{3}\!q\,
  \big(\alpha\frac{p^{2}}{2}-\alpha V_{TF}^{N,Z}(q)\big)\\
  &=-\frac{\alpha}{(2\pi)^{3}}
  \iint\limits_{|q|>\frac{1}{4}\alpha^{t}}
  d^{3}\!p\,d^{3}\!q\,
  \Big[\frac{p^{2}}{2}-\big(\frac{Z}{|q|}
  -\rho_{TF}^{N,Z}*|q|^{-1}-\mu(N)\big)\Big]_{-} \\
  &\geq-\frac{\alpha}{(2\pi)^{3}}
  \iint d^{3}\!p\,d^{3}\!q\,
  \Big[\frac{p^{2}}{2}-\big(\frac{Z}{|q|}
  -\rho_{TF}^{N,Z}*|q|^{-1}-\mu(N)\big)\Big]_{-}.
\end{align*}
Let us now look at the last term in~(\ref{eq:clas_E}).
Again using that $ V_{TF}^{N,Z}(x)\leq Z/|x|$, we have that
\begin{align*}
  \iint\limits_{|q|>\frac{1}{4}\alpha^{t};
  \ \alpha\frac{p^{2}}{2}<\alpha V_{TF}^{N,Z}(q)}
  \!\!\!\!\!\!\!\!\!\!\!\!\!\!\!
  \frac{p^{4}}{8}\,&d^{3}\!p\,d^{3}\!q\,
  \leq \!\!\!\!\!\!
  \iint\limits_{|q|>\frac{1}{4}\alpha^{t};
  \ \alpha\frac{p^{2}}{2}<\delta/|q|}
  \!\!\!\!\!\!\!\!\!\!\!\!\!
  \frac{p^{4}}{8}\,d^{3}\!p\,d^{3}\!q\,\\
  &=
  (4\pi)^{2}\int_{\frac{1}{4}\alpha^{t}}^{\infty}d|q|\,\left(|q|^{2}
  \int_{0}^{\sqrt{2Z/|q|}}\frac{|p|^{4}}{8}|p|^{2}\,d|p|\right)\\
  &=2\pi^{2}\int_{\frac{1}{4}\alpha^{t}}^{\infty}d|q|\,
  \left(|q|^{2}\Big[
  t^{7}\!/7\Big]_{0}^{\sqrt{2Z/|q|}}\right)\\
  &=\frac{2\pi^{2}(2Z)^{7/2}}{7}\int_{\frac{1}{4}\alpha^{t}}^{\infty}
  |q|^{-3/2}\,d|q|
  =\frac{8\pi^{2}(2Z)^{7/2}}{7}\alpha^{-t/2}.
\end{align*}
Using this, we then get the following
\begin{align*}
  &\frac{1}{(2\pi)^{3}}
  \!\!\!\!\!\!\!\!\!\!\!\!\!\!\!\!
  \iint\limits_{|q|>\frac{1}{4}\alpha^{t};
  \ \alpha\frac{p^{2}}{2}<\alpha V_{TF}^{N,Z}(q)}
  \!\!\!\!\!\!\!\!\!\!\!\!\!\!\!\!
  d^{3}\!p\,d^{3}\!q\,
  \big(\tilde T(p)-\alpha V_{TF}^{N,Z}(q)\big)\\
  &\geq 
  -\frac{\alpha}{(2\pi)^{3}}\iint
  d^{3}\!p\,d^{3}\!q\,
  \Big[ \frac{p^{2}}{2}-\big(\frac{Z}{|q|}
  -\rho_{TF}^{N,Z}*|q|^{-1}-\mu(N)\Big]_{-} \\
  &- \alpha^{(6-t)/2}\,\,\frac{(2Z)^{7/2}}{7\pi}.
\end{align*}
Hence, since $\delta=Z\alpha$ is fixed and $t<2/3$, we have
\begin{equation*}
 \alpha^{(6-t)/2}\,\,\frac{(2Z)^{7/2}}{7\pi}
 =\frac{8\sqrt{2}}{7\pi}\alpha^{-(1+t)/2}\delta^{7/2}
=o(\alpha^{-4/3})\quad,\quad\alpha\to0.
\end{equation*}

Summing up, we have now proved that for
$\psi\in{\mathcal H}_F=\bigwedge^N L^2(\R^3,\C^2)\,$:
\begin{align}
\label{eq:before_TF}
 \langle\psi,H\psi\rangle&\geq
 -\frac{\alpha}{(2\pi)^{3}}\iint d^{3}\!p\,d^{3}\!q\,
 \Big[\frac{p^{2}}{2}
 -(\frac{Z}{|q|}-\rho_{TF}^{N,Z}*|q|^{-1}-\mu(N))\Big]_{-} \notag\\
 &\quad-\frac{\alpha}{2}\iint
 \frac{\rho_{TF}^{N,Z}(x)\rho_{TF}^{N,Z}(y)}{|x-y|}\,d^{3}\!xd^{3}\!y
 -\alpha\mu(N)N
 -  o(\alpha^{-4/3}).
\end{align}
Integrating firstly in $p$ in the first integral in~\eqref{eq:before_TF},
we get, for each $q$ fixed:
\begin{align}
\label{eq:5/2-int}
  \int &d^{3}\!p
  \Big[\frac{p^{2}}{2}
  -(\frac{Z}{|q|}-\rho_{TF}^{N,Z}*|q|^{-1}-\mu(N))\Big]_{-}\\
  &=\!\!\!\!
  \int\limits_{\frac{p^{2}}{2}<V_{TF}^{N,Z}}
  \!\!\!\!\!
  \Big(\frac{p^{2}}{2}-(\frac{Z}{|q|}-\rho_{TF}^{N,Z}*|q|^{-1}-\mu(N))\Big)
  d^{3}\!p \notag
  =-\frac{16\sqrt{2}\pi}{15}
  \Big[V_{TF}^{N,Z}(q)\Big]_{+}^{5/2}.\notag
\end{align}
The $[\,\cdots]_{+}$, since, if the term in brackets is
negative, the integrand in~(\ref{eq:5/2-int}) will be zero.

Now, because $\rho_{TF}^{N,Z}$ satifies the equation~(\ref{eq:TF-eq}), we
get, that 
\begin{align*}
  \Big[\frac{Z}{|q|}-\rho_{TF}^{N,Z}*|q|^{-1}-\mu(N)\Big]_{+}^{5/2}
  &=\gamma^{5/2}\rho_{TF}^{N,Z}(q)^{5/3} \\
  &=\gamma^{3/2}\rho_{TF}^{N,Z}(q)
  \Big[\frac{Z}{|q|}-\rho_{TF}^{N,Z}*|q|^{-1}-\mu(N)\Big].
\end{align*}
In the last equation, no $[\,\cdots]_{+}$ is needed, since, if the last
term is negative, $\rho_{TF}^{N,Z}$ is zero, because of~(\ref{eq:TF-eq}).
In this way, by the above and by~(\ref{eq:mu-rho-N}):
\begin{align*}
 &-\frac{\alpha}{(2\pi)^{3}}\iint d^{3}\!p\,d^{3}\!q\,
 \Big[\frac{p^{2}}{2}
 -(\frac{Z}{|q|}-\rho_{TF}^{N,Z}*|q|^{-1}-\mu(N))\Big]_{-} \notag\\
 &\qquad\qquad\qquad-\frac{\alpha}{2}\iint
 \frac{\rho_{TF}^{N,Z}(x)\rho_{TF}^{N,Z}(y)}{|x-y|}\,d^{3}\!xd^{3}\!y
 -\alpha\mu(N)N\\
 &=\alpha\frac{3}{5}\gamma\int\rho_{TF}^{N,Z}(q)^{5/3}\,d^{3}\!q
 -\alpha\int\rho_{TF}^{N,Z}(q)\frac{Z}{|q|}\,d^{3}\!q\\
 &\qquad\qquad\qquad+\alpha\int\rho_{TF}^{N}(q)\,\rho_{TF}^{N,Z}*|q|^{-1}\,d^{3}\!q
 +\alpha\mu(N) \int\rho_{TF}^{N,Z}(q)\,d^{3}\!q\\
 &\qquad\qquad\qquad-\frac{\alpha}{2}\iint
 \frac{\rho_{TF}^{N,Z}(x)\rho_{TF}^{N,Z}(y)}{|x-y|}\,d^{3}\!xd^{3}\!y
 -\alpha\mu(N)N\\
 &=\alpha\left(\frac{3}{5}\gamma\int\rho_{TF}^{N,Z}(x)^{5/3}\,d^{3}\!x
 -\int\rho_{TF}^{N,Z}(x)\frac{Z}{|x|}\,d^{3}\!x
 +\frac{1}{2}\iint
 \frac{\rho_{TF}^{N,Z}(x)\rho_{TF}^{N,Z}(y)}{|x-y|} 
 \,d^{3}\!x\,d^{3}\!y\right)\\
 &=\alpha\,\mathcal{E}_{TF}(N,Z).
\end{align*}
Since $H_{\text{\it{rel}}}=\alpha^{-1}H$, and $Z=\delta\alpha^{-1}$,
with $\delta$ fixed, $0\leq\delta\leq 2/\pi$,
this shows, that for all 
$\psi\in{\mathcal H}_F=\bigwedge^N L^2(\R^3,\C^2)\,$:
\begin{align*}
  \langle\psi,H_{\text{\it{rel}}}\psi\rangle\geq
  {}-C_{TF}Z^{7/3}- o(Z^{7/3})\quad,\quad Z\to\infty,
\end{align*}
because of the scaling~\eqref{eq:scale_E}.
This ends the proof of Theorem~\ref{thm:TF}.
\qed

\appendix

\section{A formula for the kinetic energy}
\label{appendixA}

In this appendix we shall prove the localisation-formula
(\ref{eq:loc_formula}) for the operator \(\sqrt{p^{2}+\alpha^{-2}}\)
(which is the equivalent of the IMS Localisation Formula for the
Laplace operator \(-\Delta\), see Cycon, Froese, Kirsch and
Simon~\cite[Theorem 3.2]{CFKS}). Let firstly \(K_{2}\) be a modified
Bessel-function of second order, defined on \((0,\infty)\) by
\begin{align}
  K_{2}(t)=\frac{1}{2}\int_{0}^{\infty}xe^{-\frac{1}{2}t(x+x^{-1})}\,dx.
  \nonumber
\end{align}
It is easily seen that \(K_{2}\) is well-defined, decreasing and
differentiable. Other properties of \(K_{2}\) will be derived
later. Let then \(\chi_{j}, j=1,\ldots,k\) be smooth positive
functions on \(\R^{3}\), such that \(\sum_{j}\chi_{j}^{2}(x)=1\) for
all \(x\) in \(\R^{3}\) and define on \(L^{2}(\R^{3})\) the bounded
operator \(L^{(\alpha)}\) by the kernel
\begin{align}
  L^{(\alpha)}(x,y)=\frac{\alpha^{-2}}{(2\pi)^{2}}\,
  \frac{K_{2}(\alpha^{-1}|x-y|)}{|x-y|^{2}}
  \sum_{j=1}^{k}(\chi_{j}(x)-\chi_{j}(y))^{2}.
  \nonumber
\end{align}
Then for \(f\in\mathcal{S}(\R^{3})\) one has the formula:
\begin{align}
  \label{eq:a_loc_form}
  (f,\sqrt{p^{2}+\alpha^{-2}}f)
  =\sum_{j=1}^{k}(f,\chi_{j}\sqrt{p^{2}+\alpha^{-2}}\chi_{j}f)
  -(f,L^{(\alpha)}f).
\end{align}
The proof of the localisation formula (\ref{eq:a_loc_form}) will be a
consequence of the following formula:
\begin{lemma}
  For \(f\in\mathcal{S}(\R^{3})\) ,
  \begin{align}
    \label{eq:energy_form}
    \big(&f,(\sqrt{p^{2}+\alpha^{-2}}-\alpha^{-1})f\big)
   =\frac{\alpha^{-2}}{(2\pi)^{2}}
    \iint
    |f(x)-f(y)|^{2}\frac{K_{2}(\alpha^{-1}|x-y|)}{|x-y|^{2}}
    \,d^{3}\!x\,d^{3}\!y. 
  \end{align}
\end{lemma}
\begin{proof}
  Let \(\hat{f}\) be the Fourier transform of \(f\). 
  Note that by dominated convergence in momentum space, we have
  \begin{align}
    (f,\sqrt{p^{2}+\alpha^{-2}}f)
    =\lim_{t\searrow0}\frac{1}{t}\big\{(f,f)-
    (f,e^{-t\sqrt{p^{2}+\alpha^{-2}}}f)\big\}.
    \nonumber
  \end{align}
  To calculate the integral kernel
  \(\exp[-t\sqrt{p^{2}+\alpha^{-2}}](x,y)\), expand the Fourier
  transforms:
  \begin{align}
    (&f,e^{-t\sqrt{p^{2}+\alpha^{-2}}}f)=\int |\hat{f}(p)|^{2}
    e^{-t\sqrt{p^{2}+\alpha^{-2}}}\,d^{3}\!p
    \nonumber\\
    &=\frac{1}{(2\pi)^{3}}\iint \overline{f(x)}f(y)
    \bigg(\int e^{-t\sqrt{p^{2}+\alpha^{-2}}}e^{i(x-y)\cdot
      p}\,d^{3}\!p 
    \bigg)\,d^{3}\!x\,d^{3}\!y.
    \nonumber
  \end{align}
  This is justified by the fact that
  \(f\in\mathcal{S}(\R^{3})\). Now, for \(x\), \(y\) fixed, choose
  polar coordinates \((|p|,\theta,\phi)\), for \(p\) such that
  \((x-y)\cdot p=-|p|\,|x-y|\cos\theta\). Then
  \begin{align}
    &\int e^{-t\sqrt{p^{2}+\alpha^{-2}}}e^{i(x-y)\cdot
      p}\,d^{3}\!p 
    \nonumber\\
    &=\int_{0}^{\infty}\int_{0}^{2\pi}\int_{0}^{\pi}
    e^{-t\sqrt{p^{2}+\alpha^{-2}}}e^{-i|p|\,|x-y|\cos\theta}
    \sin\theta\,d\theta\,d\phi\,|p|^{2}\,d|p|
    \nonumber\\
    &=2\pi\int_{0}^{\infty}|p|^{2}e^{-t\sqrt{p^{2}+\alpha^{-2}}}
    \bigg(\int_{-1}^{1}e^{i|p|\,|x-y|u}\,du\bigg)\,d|p|
    \quad,\quad u=-\cos\theta
    \nonumber\\
    &=\frac{4\pi}{|x-y|}\int_{0}^{\infty}
    |p|e^{-t\sqrt{p^{2}+\alpha^{-2}}}\sin(|p|\,|x-y|)\,d|p|\
    \nonumber\\
    &=\frac{4\pi}{|x-y|}t\alpha^{-2}|x-y|\big(|x-y|^{2}+t^{2}\big)^{-1}
    K_{2}\big[\alpha^{-1}(|x-y|^{2}+t^{2})^{1/2}\big]
    \nonumber
  \end{align}
  where the last equality is given in Erdelyi, Magnus, Oberhettinger
  and Tricomi~\cite[p. 75, 2.4 (35)]{erdelyi}. In this way,
  \begin{align}
    \label{eq:time_evol}
    (f,&e^{-t\sqrt{p^{2}+\alpha^{-2}}}f)
    =\frac{t\alpha^{-2}}{2\pi^{2}}\iint\overline{f(x)}f(y)
    \frac{K_{2}\big[\alpha^{-1}(|x-y|^{2}+t^{2})^{1/2}\big]}{|x-y|^{2}+t^{2}} 
    \,d^{3}\!x\,d^{3}\!y.
  \end{align}
  Now, letting \(F_{t}(p)=e^{-t\sqrt{p^{2}+\alpha^{-2}}}\), the above
  shows that
  \begin{align}
    \check{F}_{t}(x)&=\frac{1}{(2\pi)^{3/2}}\int F_{t}(p)e^{ix\cdot
    p}\,d^{3}\!p
    =\sqrt{\frac{2}{\pi}}\,t\alpha^{-2}\,
    \frac{K_{2}\big[\alpha^{-1}(|x|^{2}+t^{2})^{1/2}\big]}{|x|^{2}+t^{2}},
    \nonumber
  \end{align}
  and therefore, for all \(y\in\R^{3}\):
  \begin{align}
    \label{eq:Fourier_at_0}
    \frac{t\alpha^{-2}}{2\pi^{2}}\,
    \int
    \frac{K_{2}\big[\alpha^{-1}(|x-y|^{2}+t^{2})^{1/2}\big]}{|x-y|^{2}+t^{2}} 
    \,d^{3}\!x
    =F_{t}(0)=e^{-t\alpha^{-1}}.
  \end{align}
  Hence we get, using (\ref{eq:time_evol}) and
  (\ref{eq:Fourier_at_0}), which are both symmetric in \(x\) and
  \(y\), that
  \begin{align}
    \frac{1}{t}\big\{(f&,f)-(f,e^{-t\sqrt{p^{2}+\alpha^{-2}}}f)\big\}
    \nonumber\\
    &=\frac{1}{t}\big\{(f,f)-(f,e^{-t\alpha^{-1}}f)\big\}
    +\frac{1}{t}\big\{(f,e^{-t\alpha^{-1}}f)
    -(f,e^{-t\sqrt{p^{2}+\alpha^{-2}}}f)\big\}
    \nonumber\\
    &={} - {}
    \frac{e^{-t\alpha^{-1}}-e^{-0\cdot\alpha^{-1}}}{t-0}\;(f,f)
    \nonumber\\
    &\quad+\frac{1}{t}\Big\{
    \int
    \frac{1}{2}\Big(\big(|f(x)|^{2}+|f(y)|^{2}\big)
    -\overline{f(x)}f(y)-\overline{f(y)}f(x)\Big)
    \nonumber\\
    &\quad\quad\quad\quad
    \times\frac{t\alpha^{-2}}{2\pi^{2}}
    \frac{K_{2}\big[\alpha^{-1}(|x-y|^{2}+t^{2})^{1/2}\big]}{|x-y|^{2}+t^{2}} 
    \,d^{3}\!x\,d^{3}\!y\Big\}.
    \nonumber
  \end{align}
  Cancelling \(t\) and noting that 
  \begin{align}
    \lim_{t\searrow0}\;\frac{e^{-t\alpha^{-1}}-e^{-0\cdot\alpha^{-1}}}{t-0}
    =\left.\frac{d}{dt}
    \big(e^{-t\alpha^{-1}}\big)\right|_{t=0}
    =-\alpha^{-1},
    \nonumber
  \end{align}
  we get that
  \begin{align}
    \lim_{t\searrow0}\frac{1}{t}\big\{&(f,f)-
    (f,e^{-t\sqrt{p^{2}+\alpha^{-2}}}f)\big\}
    \nonumber\\
    &=\alpha^{-1}+\frac{\alpha^{-2}}{(2\pi)^{2}}\iint|f(x)-f(y)|^{2}
    \frac{K_{2}(\alpha^{-1}|x-y|)}{|x-y|^{2}}\,d^{3}\!x\,d^{3}\!y.
    \nonumber
  \end{align}
  This proves the lemma.
\end{proof}
Now, to prove the formula (\ref{eq:a_loc_form}), we simply use the
fact that \(\sum_{j}\chi_{j}^{2}(x)=1\) for all \(x\) in \(\R^{3}\):
\begin{align}
  &\sum_{j=1}^{k}|\chi_{j}(x)f(x)-\chi_{j}(y)f(y)|^{2}
  \nonumber\\
  &=|f(x)|^{2}+|f(y)|^{2}-\sum_{j=1}^{k}
  \chi_{j}(x)\chi_{j}(y)\big(\overline{f(y)}f(x)+\overline{f(x)}f(y)\big)
  \nonumber\\
  &=|f(x)-f(y)|^{2}+\sum_{j=1}^{k}\chi_{j}(x)
  \big(\overline{f(y)}f(x)+\overline{f(x)}f(y)\big)
  \big(\chi_{j}(x)-\chi_{j}(y)\big).
  \nonumber 
\end{align}
Note that \(\chi_{j}f\in\mathcal{S}(\R^{3})\), since \(\chi_{j}\) is
smooth and bounded, so that using the formula (\ref{eq:energy_form}):
\begin{align}
  \label{eq:way_too_long}
  \sum_{j=1}^{k}&(f,\chi_{j}(\sqrt{p^{2}+\alpha^{-2}}-\alpha^{-1})\chi_{j}f)
  =\sum_{j=1}^{k}(\chi_{j}f,(\sqrt{p^{2}+\alpha^{-2}}-\alpha^{-1})\chi_{j}f)
  \nonumber\\
  &=\frac{\alpha^{-2}}{(2\pi)^{2}}
  \iint\sum_{j=1}^{k}|\chi_{j}(x)f(x)-\chi_{j}(y)f(y)|^{2}\,
  \frac{K_{2}(\alpha^{-1}|x-y|)}{|x-y|^{2}}\,d^{3}\!x\,d^{3}\!y
  \nonumber\\
  &=\frac{\alpha^{-2}}{(2\pi)^{2}}
  \iint\Big\{|f(x)-f(y)|^{2}+\sum_{j=1}^{k}\chi_{j}(x)\big(\overline{f(y)} 
  f(x)
  \nonumber\\
  &\qquad\qquad\quad
  +\overline{f(x)}f(y)\big)\big(\chi_{j}(x)-\chi_{j}(y)\big)\Big\}
  \frac{K_{2}(\alpha^{-1}|x-y|)}{|x-y|^{2}}\,d^{3}\!x\,d^{3}\!y.
\end{align}
Using now that
\begin{align}
  \iint&\chi_{j}(x)\overline{f(y)}f(x)\big(\chi_{j}(x)-\chi_{j}(y)\big)
  \frac{K_{2}(\alpha^{-1}|x-y|)}{|x-y|^{2}}\,d^{3}\!x\,d^{3}\!y
  \nonumber\\
  &=-\iint\chi_{j}(y)\overline{f(x)}f(y)\big(\chi_{j}(x)-\chi_{j}(y)\big)
  \frac{K_{2}(\alpha^{-1}|x-y|)}{|x-y|^{2}}\,d^{3}\!x\,d^{3}\!y
  \nonumber
\end{align}
simply by interchanging \(x\) and \(y\), we finally get from 
(\ref{eq:way_too_long}) that
\begin{align}
  \sum_{j=1}^{k}&(f,\chi_{j}\sqrt{p^{2}+\alpha^{-2}}\chi_{j}f)
  =\frac{\alpha^{-2}}{(2\pi)^{2}}
  \iint|f(x)-f(y)|^{2}\frac{K_{2}(\alpha^{-1}|x-y|)}{|x-y|^{2}}
  \,d^{3}\!x\,d^{3}\!y
  \nonumber\\
  &+\frac{\alpha^{-2}}{(2\pi)^{2}}\iint\overline{f(x)}f(y)
  \sum_{j=1}^{k}\big(\chi_{j}(x)-\chi_{j}(y)\big)^{2}
  \frac{K_{2}(\alpha^{-1}|x-y|)}{|x-y|^{2}}
  \,d^{3}\!x\,d^{3}\!y
  \nonumber
\end{align}
which, using (\ref{eq:energy_form}), proves the formula
(\ref{eq:a_loc_form}). 
\hfill\qed

We now derive two facts about the function \(K_{2}\):
\begin{align}
  \label{eq:K_{2}-int}
  \int_{0}^{\infty}&t^{2}K_{2}(t)\,dt=\frac{3\pi}{2},
  \\
  \label{eq:K_{2}-est}
  K_{2}(t)\leq4\sqrt{\frac{\pi}{2t}}\,e^{-t}\,
  \big(1&+\frac{1}{2t}+\frac{1}{(2t)^{2}}\big)\quad\text{for all } t\in\R_{+}  
\end{align}

The proof of (\ref{eq:K_{2}-int}) is straightforward by using the
definition of \(K_{2}\):
\begin{align}
  \int_{0}^{\infty}t^{2}K_{2}(t)\,dt
  &=\int_{0}^{\infty}t^{2}\Big(\frac{1}{2}
  \int_{0}^{\infty}xe^{-\frac{1}{2}(x+x^{-1})}\,dx\Big)\,dt
  \nonumber\\
  &=\frac{1}{2}\int_{0}^{\infty}x\Big(\int_{0}^{\infty}t^{2}
  e^{-\frac{1}{2}(x+x^{-1})}\,dt\Big)\,dx
  \nonumber
\end{align}
where the interchanging of the order of integration is allowed by
Tonel\-li's theorem. By applying partial integration three times, 
\begin{align}
  \int_{0}^{\infty}t^{2}
  e^{-\frac{1}{2}(x+x^{-1})}\,dt=\frac{16}{(x+x^{-1})},
  \nonumber
\end{align}
and so
\begin{align}
  \int_{0}^{\infty}t^{2}K_{2}(t)\,dt
  &=
  \frac{1}{2}\int_{0}^{\infty}\frac{16x}{(x+x^{-1})}\,dx
  \nonumber
  =4\int_{-\infty}^{\infty}\frac{x^{4}}{(x^{2}+1)^{3}}\,dx
  \nonumber\\
  &=4\cdot2\pi i\,\Res\Big(\frac{z^{4}}{(z^{2}+1)^{3}},i\Big)
  =8\pi i\frac{6}{32i}=\frac{3\pi}{2}.
  \nonumber
\end{align}

For the estimate (\ref{eq:K_{2}-est}), we need to rewrite
\(K_{2}\). This is done following Gray and Mathews~\cite[pp.\
50]{GrMa}.
\begin{obs}
  \begin{align}
    \label{eq:rewrite_K_2}
    K_{2}(t)=\sqrt{\frac{\pi}{2t}}
    \,\frac{1}{\Gamma(\frac{5}{2})}\,
    e^{-t}\int_{0}^{\infty}e^{-\xi}\xi^{3/2}\big(1+\frac{\xi}{2t}\big)^{3/2}  
    \,d\xi.
  \end{align}
\end{obs}
To prove the observation, we start on the right-hand-side of
(\ref{eq:rewrite_K_2}). Setting \(t+\xi=\sqrt{t^{2}+\eta}\), one gets, 
since then \(\eta=\xi^{2}+2t\xi\), that
\begin{align}
  \text{RHS} \;(\ref{eq:rewrite_K_2})
  =\sqrt{\frac{\pi}{2t}}\,
    \frac{1}{\Gamma(\frac{5}{2})}
  \int_{0}^{\infty}e^{-\sqrt{t^{2}+\eta}}\big(\frac{\eta}{2t}\big)^{3/2}
  \frac{d\eta}{2\sqrt{t^{2}+\eta}}.
  \nonumber
\end{align}
Using the formula 
\begin{align}
  \int_{0}^{\infty}
  e^{-(a^{2}\xi^{2}+b^{2}/\xi^{2})}\,d\xi=\frac{\sqrt{\pi}}{2a}\,e^{-2ab}
  \nonumber
\end{align}
(which holds since both sides satisfy the differential equation 
\(df/db=-2af, f(b=0)=\sqrt{\pi}/2a\))
with \(a=\sqrt{t^{2}+\eta}\), \(b=1/2\), we arrive at
\begin{align}
  \text{RHS} \;(\ref{eq:rewrite_K_2})
  &=\frac{1}{\Gamma(\frac{5}{2})(2t)^{2}}
  \int_{0}^{\infty}\eta^{3/2}\Big(
  \int_{0}^{\infty}e^{-\big((t^{2}+\eta)\xi^{2}+1/(2\xi)^{2}\big)}
  \,d\xi\Big)\,d\eta
  \nonumber\\
  &=\frac{1}{\Gamma(\frac{5}{2})(2t)^{2}}\int_{0}^{\infty}
  e^{-(t^{2}\xi^{2}+1/(2\xi)^{2})}\Big(\int_{0}^{\infty}
  e^{-\eta\xi^{2}}\eta^{3/2}\,d\eta\Big)\,d\xi
  \nonumber\\
  &=\frac{1}{(2t)^{2}}
  \int_{0}^{\infty}e^{-(t^{2}\xi^{2}+1/(2\xi)^{2})}\xi^{-5}\,d\xi
  \nonumber
\end{align}
since one has the formula
\begin{align}
  \int_{0}^{\infty}e^{-\eta\xi^{2}}\eta^{3/2}\,d\eta
  =\xi^{-5}\Gamma(\tfrac{5}{2}) .
  \nonumber
\end{align}
Making the change of variables \(x=\frac{1}{2t\xi^{2}}\), we finally get
\begin{align}
  \text{RHS} \;(\ref{eq:rewrite_K_2})=
  \frac{1}{2}\int_{0}^{\infty}
  xe^{-\frac{1}{2}t(x+x^{-1})}\,dx=K_{2}(t).
  \nonumber
\end{align}
Now, to prove the estimate (\ref{eq:K_{2}-est}), use the Tayloer
expansion (\ref{eq:Taylor}) on the integrand in
(\ref{eq:rewrite_K_2}), to get
\begin{align}
  K_{2}(t)
  &\leq\sqrt{\frac{\pi}{2t}}\,\frac{1}{\Gamma(\frac{5}{2})}\,
  e^{-t}\int_{0}^{\infty}e^{-\xi}\xi^{3/2}
  \big(1+\frac{3}{4t}\xi+\frac{3}{32t^{2}}\xi^{2}\big)\,d\xi
  \nonumber\\
  &=\sqrt{\frac{\pi}{2t}}\,\frac{1}{\Gamma(\frac{5}{2})}\,
  e^{-t}\bigg(\int_{0}^{\infty}e^{-\xi}\xi^{3/2}\,d\xi
  \nonumber\\
  &\qquad\qquad\qquad\qquad
  +\frac{3}{4t}\int_{0}^{\infty}e^{-\xi}\xi^{5/2}\,d\xi
  +\frac{3}{32t^{2}}\int_{0}^{\infty}e^{-\xi}\xi^{7/2}\,d\xi
  \bigg)
  \nonumber\\
  &=\sqrt{\frac{\pi}{2t}}\,\frac{1}{\Gamma(\frac{5}{2})}\,
  e^{-t}\Big(\Gamma(\tfrac{5}{2})+\frac{3}{4t}\Gamma(\tfrac{7}{2})
  +\frac{3}{32t^{2}}\Gamma(\tfrac{9}{2})\Big)
  \nonumber\\
  &=\sqrt{\frac{\pi}{2t}}\,
  e^{-t}\big(1+\frac{15}{8t}+\frac{105}{128t^{2}}\big)
  \leq4\sqrt{\frac{\pi}{2t}}\,
  e^{-t}\big(1+\frac{1}{2t}+\frac{1}{(2t)^{2}}\big).
  \nonumber
\end{align}

\section{Introducing coherent states}
\label{appendixB}

In this section we will introduce coherent states and prove the
formulae
in section \ref{outer}. The error introduced by using
coherent states will also be estimated here.

\begin{lemma}
Let $g\in C_{0}^{\infty}(\R^{3})$ be spherically symmetric,
non-negative, supported in the unit ball and such that 
$\|g\|_{2}=1$, and let $g^{p,q}(x)=g(x-q)e^{ipx}$. Then 
\begin{align}
  (f,f)&=\frac{1}{(2\pi)^{3}}\iint 
  d^{3}\!p\,d^{3}\!q\,(f,g^{p,q})(g^{p,q},f) \notag\\
  (f,(V*|g|^{2})f) &= 
  \frac{1}{(2\pi)^{3}}\iint d^{3}\!p\,d^{3}\!q\,V(q)
  (f,g^{p,q})(g^{p,q},f)\notag\\
  (f,\sqrt{p^{2}+\alpha^{-2}}f) &\geq 
  \frac{1}{(2\pi)^{3}}\iint d^{3}\!p\,d^{3}\!q\,\sqrt{p^{2}+\alpha^{-2}}\,
  (f,g^{p,q})(g^{p,q},f)\notag\\
  &\quad-3\alpha\,\|\nabla g\|_{\infty}^{2}\Vol(\supp\ g)\,\|f\|_{2}^{2}.
\end{align}
\end{lemma}

\begin{proof}
The idea of the above formulae is to write the identity  and
other operators on $L^{2}(\R^{3})$ as superpositions of the
one-rank operators $\pi_{pq}=(\ \ ,g^{p,q})g^{p,q}$. 
To prove the above formulae, start with the right-hand-side
of the second formula (the proof of the first formula is
similar, just more simple):
\begin{align}
  \label{eq:Fou_trick}
  &\frac{1}{(2\pi)^{3}}\iint d^{3}\!p\,d^{3}\!q\,V(q)
  (f,g^{p,q})(g^{p,q},f)\\
  &=\frac{1}{(2\pi)^{3}}\iint d^{3}\!p\,d^{3}\!q\,V(q)\Big[\overline{
  \int f(y)\overline{g(y-q)}e^{-ipy}\,d^{3}\!y}\Big] 
  \Big[\int f(x)\overline{g(x-q)}e^{-ipx}\,d^{3}\!x\Big]
  \nonumber
\end{align}
Notice, that the function in the last brackets is $(2\pi)^{3/2}$ times
the Fourier-transform of the function
$F_{q}(x)=f(x)\overline{g(x-q)}$.
In this way we get, by Parseval's formula:
\begin{align*}
 \eqref{eq:Fou_trick} & =
 \iint d^{3}\!p\,d^{3}\!q\,V(q) \,|\hat F_{q}(p)|^{2}
 =\int d^{3}\!q\,V(q) \,\|\hat F_{q}\|_{2}^{2}
 = \int d^{3}\!q\,V(q) \,\|F_{q}\|_{2}^{2}
 \\&
 = \int d^{3}\!q\, V(q) \Big( \int|f(x)|^{2}\,|g(x-q)|^{2}\,d^{3}\!x\Big)\\
 &= \int d^{3}\!x\, |f(x)|^{2}\Big(\int V(q)\,|g(x-q)|^{2}\,d^{3}\!q\Big)
 = (f,(V*|g|^{2})f) .
\end{align*}
This proves the second (and the first) formula. 

To prove the formula for the operator $\sqrt{p^{2}+\alpha^{-2}}$,
note that 
\begin{equation*}
 \int g(x-q)^{2}\,d^{3}\!q=1 \text{ for all }x \text{ in } \R^{3},
\end{equation*}
so that, by the symmetry of the operator $\sqrt{p^{2}+\alpha^{-2}}$:
\begin{align}
\label{eq:coh_1}
 (f,\sqrt{p^{2}+\alpha^{-2}}f)&=\frac{1}{2}
 \iint\overline{f(x)}g(x-q)^{2}\big(\sqrt{p^{2}+\alpha^{-2}}f\big)(x)
 \,d^{3}\!q\,d^{3}\!x
 \notag\\&
 \quad+\frac{1}{2}\iint\overline{\big(\sqrt{p^{2}+\alpha^{-2}}f\big)(x)}\,
 g(x-q)^{2}f(x)
 \,d^{3}\!q\,d^{3}\!x
 \notag\\&
  =\frac{1}{2}
 \iint\overline{f(x)}g_{q}(x)^{2}\big(\sqrt{p^{2}+\alpha^{-2}}f\big)(x)
 \,d^{3}\!q\,d^{3}\!x\notag\\
 &\quad+\frac{1}{2}\iint\overline{f(x)}
 \Big(\sqrt{p^{2}+\alpha^{-2}}\big(g_{q}{}^{2}f\big)\Big)(x)
 \,d^{3}\!q\,d^{3}\!x.
\end{align}
Here, $g_{q}(x)=g(x-q)$. Remembering that $g_{q}(x)^{2}$ is reel and
letting $g_{q}{}^{2}$ denote the
multiplication operator defined by this function, we have
\begin{align}
  \label{eq:coh_2}
  &\eqref{eq:coh_1} =\iint\overline{(g_{q}f)(x)}
  \big[\sqrt{p^{2}+\alpha^{-2}}(g_{q}f)\big](x)
  \,d^{3}\!q\,d^{3}\!x\\
   &  +\frac{1}{2}\iint\overline{f(x)}\Big[\big(
  g_{q}{}^{2}\sqrt{p^{2}+\alpha^{-2}}
  +\sqrt{p^{2}+\alpha^{-2}}g_{q}{}^{2}- 
  2g_{q}\sqrt{p^{2}+\alpha^{-2}}g_{q}\big)f\Big](x)
  \,d^{3}\!q\,d^{3}\!x
  \notag\\
  &=\frac{1}{2}\iint\overline{f(x)}\big(L_{q}f\big)(x)
  \,d^{3}\!q\,d^{3}\!x
  \nonumber\\
  &+\iint
  \Big(\int\sqrt{p^{2}+\alpha^{-2}}
  \Big(\int e^{-ipy}g_{q}(y)f(y)\,d^{3}\!y\Big)e^{ipx}\,d^{3}\!p\Big)
  g_{q}(x)\overline{f(x)},
  \,d^{3}\!q\,d^{3}\!x\nonumber
\end{align}
where
\begin{align}
\label{eq:L_q}
 &\big(L_{q}f\big)(x)=\\
 &\qquad\int\Big\{\int\big[g_{q}(y)^{2}+g_{q}(x)^{2}
 -2g_{q}(x)g_{q}(y)\big]\sqrt{p^{2}+\alpha^{-2}}
 e^{ip(x-y)}\,d^{3}\!p\Big\}
 f(y)\,d^{3}\!y.\nonumber
\end{align}
The second term in~(\ref{eq:coh_2}) is equal to
\begin{align*}
 &\iint\,d^{3}\!p\,d^{3}\!q\,\sqrt{p^{2}+\alpha^{-2}}
 \left(\int\overline{f(x)}g_{q}(x)e^{ipx}\,d^{3}\!x\right)
 \left(\int f(y)g_{q}(y)e^{-ipy}\,d^{3}\!y\right)\\
 &=\iint\,d^{3}\!p\,d^{3}\!q\,\sqrt{p^{2}+\alpha^{-2}}\,
 (f,g^{p,q})(g^{p,q},f).
\end{align*}
The first term in~(\ref{eq:coh_2}) is the error, which will now
be estimated. Keeping $x$ and $y$ fixed, we have, as showed in the
proof of~(\ref{eq:energy_form}):
\begin{align*}
  L_{q}(x,y)&=\int\big[g_{q}(y)^{2}+g_{q}(x)^{2}
 -2g_{q}(x)g_{q}(y)\big]\sqrt{p^{2}+\alpha^{-2}}\,
 e^{ip(x-y)}\,d^{3}\!p\\
 &=\big[g_{q}(x)-g_{q}(y)\big]^{2}\frac{\alpha^{-2}}{4\pi^{2}}
 \frac{K_{2}(\alpha^{-1}|x-y|)}{|x-y|^{2}}.
\end{align*}
In this way, using the same ideas as in Section~\ref{loc_error}, 
we reach the estimate
\begin{align*}
  L_{q}(x,y)\leq\|\nabla
  g_{q}\|_{\infty}^{2}\,\frac{\alpha^{-2}}{4\pi^{2}}\,
  K_{2}(\alpha^{-1}|x-y|)\,
  \big(\chi_{\supp g_{q}}(x)+\chi_{\supp g_{q}}(y)\big),
\end{align*}
where $\chi_{\supp g_{q}}$ is the characteristic function of $\supp g_{q}$.
This gives us that
\begin{align*}
  \int &L_{q}(x,y)\,d^{3}\!q
  \leq
  \int \|\nabla
  g_{q}\|_{\infty}^{2}\,\frac{\alpha^{-2}}{4\pi^{2}}\,
  K_{2}(\alpha^{-1}|x-y|)\,
  \big(\chi_{\supp g_{q}}(x)+\chi{\supp g_{q}}(y)\big)
  \,d^{3}\!q\\
  &= 2\,\|\nabla
  g\|_{\infty}^{2}\,\frac{\alpha^{-2}}{4\pi^{2}}\,
  K_{2}(\alpha^{-1}|x-y|)\,\Vol(\supp g).
\end{align*}
By this we finally get, by using first Cauchy-Schwartz's, then
Young's inequality, that
\begin{align*}
  \Big| &\iint\overline{f(x)}\int L_{q}(x,y)\,d^{3}\!q\,f(y)\,d^{3}\!x
  \,d^{3}\!y \Big| \\ 
  &\leq
  \iint |f(x)|\,\Big(2\,\|\nabla g\|_{\infty}^{2}
  \,\frac{\alpha^{-2}}{4\pi^{2}}\,
  K_{2}(\alpha^{-1}|x-y|)\,\Vol(\supp g)\Big)|f(y)|\,d^{3}\!x\,d^{3}\!y\\
  &\leq 
  2\,\|\nabla g\|_{\infty}^{2}\,\frac{\alpha^{-2}}{4\pi^{2}} \,
  \|f\|_{2}\,\||f|*G_{\alpha}\|_{2}\,\Vol(\supp g)
  \quad,\  G_{\alpha}(x)=K_{2}(\alpha^{-1}|x|)\\
  &\leq 
  \|\nabla g\|_{\infty}^{2}\,\frac{\alpha^{-2}}{2\pi^{2}}\,
  \|f\|_{2}^{2}\,\|G_{\alpha}\|_{1}\,\Vol(\supp g)\\
  &=\|\nabla
  g\|_{\infty}^{2}\,\frac{\alpha^{-2}}{2\pi^{2}} 
  \,6\pi^{2}\alpha^{3} \|f\|_{2}^{2}\,\Vol(\supp g)
  \qquad\text{(see~\ref{eq:K_{2}-int} for $\|G_{\alpha}\|_{1}$)} \\
  &=3\alpha\,\|\nabla g\|_{\infty}^{2}\Vol(\supp g)\,\|f\|_{2}^{2}.
\end{align*}
\end{proof}
For the case~\eqref{eq:coh_states} in Section~\ref{outer}, 
let the coherent state $g^{p,q}$
be defined from the scaled version of the function $g$ chosen 
there---that is, $g\in C_{0}^{\infty}(\R^{3})$,
spherically symmetric, non-negative and with support in the unit ball
$B(0,1)$ of $\R^{3}$. Then the coherent states are
\begin{equation*}
  g_{\alpha}^{p,q}(x)=g_{\alpha}(x-q)e^{ipx}
  =\alpha^{-3s/2}g\big(\frac{x-q}{\alpha^{s}}\big)e^{ipx}.
\end{equation*}
In this way, $\|\nabla 
g_{\alpha}\|_{\infty}^{2}=\alpha^{-5s}\|\nabla g\|_{\infty}^{2}$
and $\Vol(\supp g_{\alpha})=\frac{4\pi}{3}\alpha^{3s}$, and therefore
\begin{align*}
  (f,\sqrt{p^{2}+\alpha^{-2}}f) &\geq 
  \frac{1}{(2\pi)^{3}}\iint d^{3}\!p\,d^{3}\!q\,\sqrt{p^{2}+\alpha^{-2}}\,
  (f,g_{\alpha}^{p,q})(g_{\alpha}^{p,q},f)
  \ -\ o(\alpha^{-1/3}),
\end{align*}
since, as $s<2/3$,
\begin{equation*}
  3\alpha\alpha^{-5s}\,\|\nabla g\|_{\infty}^{2}
  \frac{4\pi}{3}\alpha^{3s}\,\|f\|_{2}^{2} = C\,\alpha^{1-2s}
  = o(\alpha^{-1/3})\ ,\ \alpha\to0.
\end{equation*}
This proves the formula~\eqref{eq:coh_states}, 
since
\begin{equation*}
 (f,f)=\frac{1}{(2\pi)^{3}}\iint
  d^{3}\!p\,d^{3}\!q\,(f,g_{\alpha}^{p,q})
  (g_{\alpha}^{p,q},f)
\end{equation*}
and $T(p)=\sqrt{p^{2}+\alpha^{-2}}-\alpha^{-1}$.

\def\cprime{$'$} \def\cprime{$'$} \def\cprime{$'$}
\providecommand{\bysame}{\leavevmode\hbox to3em{\hrulefill}\thinspace}
\providecommand{\MR}{\relax\ifhmode\unskip\space\fi MR }
\providecommand{\MRhref}[2]{%
  \href{http://www.ams.org/mathscinet-getitem?mr=#1}{#2}
}
\providecommand{\href}[2]{#2}

%Til BiBTeX-brug:
%\bibliographystyle{amsplain}
%\bibliography{local}    

\end{document}